\documentclass[a4paper,11pt]{article}

\usepackage{amssymb,amstext,amsmath,amsthm}
\usepackage{graphicx}
\usepackage{latexsym}
\usepackage{psfrag}
\usepackage{amsfonts}
\usepackage{helvet}

\usepackage{amsmath,amsxtra,amsthm,amssymb,xr}
\usepackage[all]{xy}
\usepackage[]{mathrsfs}
\usepackage{vmargin}

\newcommand{\bea}{\begin{eqnarray}}
\newcommand{\eea}{\end{eqnarray}}
\newcommand{\nn}{\nonumber}
\newcommand{\be}{\begin{equation}}
\newcommand{\ee}{\end{equation}}

\newtheorem{theo}{Theorem}
\newtheorem{lem}{Lemma}

\newcommand{\N}{\mathbb{N}}

\newcommand{\R}{\mathbb{R}}
\newcommand{\C}{\mathbb{C}}

\DeclareMathOperator{\tr}{tr}

\newcommand{\lalg}[1]{\mathfrak{#1}}  
\newcommand{\SU}{\mathrm{SU}}
\newcommand{\SO}{\mathrm{SO}}

\newcommand{\U}{\mathrm{U}}
\newcommand{\Spin}{\mathrm{Spin}}
\newcommand{\Hom}{\mathrm{Hom}}
\newcommand{\Reel}{\mathrm{Re}}
\newcommand{\su}{\lalg{su}}

\newcommand{\so}{\lalg{so}}

\newcommand{\spin}{\lalg{spin}}
\newcommand{\pole}{\mathcal N} 

\def\la{\langle}
\def\ra{\rangle}

\def\vphi{\varphi}

\setmarginsrb{2.5cm}{1cm}{2.5cm}{2cm}{12pt}{11mm}{0pt}{13mm}

\title{Asymptotic analysis of the EPRL four-simplex amplitude}
\author{John W. Barrett\footnote{john.barrett@nottingham.ac.uk}
, Richard J. Dowdall\footnote{richard.dowdall@maths.nottingham.ac.uk}
, Winston J. Fairbairn\footnote{winston.fairbairn@nottingham.ac.uk}
, \\ Henrique Gomes\footnote{henrique.gomes@maths.nottingham.ac.uk}
, Frank Hellmann\footnote{frank.hellmann@maths.nottingham.ac.uk}
\vspace{3mm}
\\ School of Mathematical Sciences \\ Nottingham University \\ University Park \\ Nottingham NG7 2RD \\ UK}

\date{}

\begin{document}

\maketitle

\vspace{-3mm}
\begin{abstract}
The semiclassical limit of a 4-simplex amplitude for a spin foam quantum gravity model with an Immirzi parameter is studied.
If the boundary state represents a non-degenerate 4-simplex geometry, the asymptotic formula contains the Regge action for general relativity. A canonical choice of phase for the boundary state is introduced and is shown to be necessary to obtain the results. 
\end{abstract}

\section{Introduction}

A key step in understanding the semiclassical regime of a spin foam model \cite{baez-2000-543,oriti-2001-64,perez-2003-20} in dimension $n$ is the analysis of the asymptotic behaviour of the $n$-simplex amplitude that defines the model. In fact, the discovery by Ponzano and Regge that the 6j symbol of recoupling theory contains the Regge action in its asymptotic behaviour established it as a model for 3D quantum gravity \cite{ponzanoregge}. Similar asymptotic analysis of the 4-dimensional models \cite{barrett-1998-39} was initially performed by Barrett, Williams and Steele \cite{barrett-1999-3,barrett-2003-20}, and formed the basis of investigations of the graviton propagator structure of these models \cite{Alesci:2007tx,Alesci:2007tg,Alesci:2008gv}. This latter analysis showed a definite incompatibility between the 10j symbol and a boundary structure given by loop quantum gravity-like geometry. Consequently a host of new 4-dimensional models were developed. The first such models, were due to Engle, Pereira and Rovelli \cite{Engle:2007qf}. Meanwhile Livine and Speziale introduced coherent states to the analysis and definition of spin foam models \cite{livine-2007-76}, and suggested a  way to construct new models in \cite{Livine:2007ya}. In parallel, Freidel and Krasnov defined and developed a model along these lines in \cite{Freidel:2007py}. A refined version of the original EPR model was published in \cite{Engle:2007wy} together with Livine. The FK model and the EPRL model depend on the Immirzi parameter $\gamma$ and are identical for $\gamma < 1$. An initial exploration of the asymptotics of the FK model for manifolds without boundary was undertaken in \cite{Conrady:2008mk}.

Our thanks are due to Carlo Rovelli, who encouraged us to study this problem, and for hosting JWB and WF in Marseille, where the methods and most of the results were presented in detail on 19th November 2008 in a five-hour seminar. The result presented in Marseille was a complete derivation of the term in the asymptotic formula which gives the cosine of the Regge action for $\gamma<1$, and is described by Alesci, Bianchi and Rovelli in \cite{Alesci:2008ff}, and also in the talk by Alesci at PI \cite{Alescitalk}.  The derivation of the phase part of this term is also confirmed by the work of Freidel and Conrady in \cite{Conrady:2009px}.

A spin foam model is a procedure to compute an amplitude from a triangulated manifold $\mathcal{T}$ with $n$-simplices $\Delta_n$ coloured by representation theory data. In four-dimensions, such an amplitude is typically of the form
\be
\mathcal{Z}(\mathcal{T}) = \sum_{\iota, \rho} \prod_{\Delta_2} f_2 (\rho)
\prod_{\Delta_3} f_3(\rho,\iota) \prod_{\Delta_4} f_4(\rho,\iota).
\ee
where $f_n$ are weights assigned to the $n$-simplices of the triangulated manifold, and $\rho$ and $\iota$ respectively denote the assignments of unitary, irreducible representations to the $2$-simplices, and intertwining operators to the $3$-simplices of $\mathcal{T}$. The model is specified by the choice of representation assignments, the vector space of intertwining operators $\iota$, and weights $f_n$.

The Euclidean EPRL is a specific spin foam model \cite{Engle:2007wy} built out of the representation theory of the spin group $G=\Spin(4)$. In this paper we analyse in detail the large spin structure of the four-simplex amplitude $f_4$ of the model. We begin by defining the EPRL model in section \ref{EPRL}. In section \ref{posing} we discuss the general well-posedness of the analysis to be performed here and discuss the geometry of the three dimensional boundary determined by the chosen intertwiners $\iota$. In section \ref{asymptotic} we state the main result of the full vertex amplitude, which we then prove in section \ref{proof}. We conclude by discussing some further aspects of the results obtained in section \ref{Conclusion}.

\section{The EPRL four-simplex amplitude}\label{EPRL}

The input data for the 4-simplex amplitude is a spin $k\in\{0,\frac12,1,\ldots\}$ for each triangle of the 4-simplex and an $\SU(2)$ intertwiner $\hat{\iota}$ for each tetrahedron. From $\hat{\iota}$, a $\Spin(4)$ intertwiner $\iota$ is constructed, and then these  $\Spin(4)$ intertwiners are glued together in the standard fashion to construct an amplitude (a complex number) for this data. The only other input required is the Immirzi parameter $\gamma$, which is a constant.

Firstly, a precise definition of $\hat{\iota}$ is required. For a given tetrahedron, one has to choose an ordering of its four faces, e.g., by numbering them with $1,2,3,4$. Then the $\SU(2)$ intertwiner $\hat{\iota}$ is an element of
$$\Hom_{\SU(2)}(\C,\bigotimes_{i=1}^4 V_{k_i}),$$
where the spaces are tensored together in the order $V_{k_1}\otimes V_{k_2}\otimes V_{k_3}\otimes
V_{k_4}$.
This ordering convention is used throughout.

Of course the spaces constructed using different orderings are easily related by an action of the permutation group. We use the {\em binor} category of representations \cite{Major:1999md,penrose,Barrett:2008wh} throughout the paper. In this category the crossing diagram is fermionic, which means that the crossing of two lines of odd spin gives a factor of $-1$. For example, the map $V_{k_1}\otimes V_{k_2}\to V_{k_2}\otimes V_{k_1}$ is  $x\otimes y\to (-1)^{4k_1k_2} y\otimes x$. Spin network diagrams in this category can be evaluated using the Kauffman bracket \cite{KauffmanLins} specialised to Kauffman's parameter
$A=-1$. The binor calculus has the convenient feature that the framing of a curve does not affect the evaluation.

The $\iota$ are constructed as follows. Let $(\pi_k,V_k)$ and
$(\pi_{(j^-,j^+)},V_{(j^-,j^+)})$ respectively denote the unitary, irreducible representations of $\SU(2)$ and
$\Spin(4) = \SU(2)_- \times \SU(2)_+$. There exists an injection
\bea
\phi: \Hom_{\SU(2)}(\C,\bigotimes_{i=1}^4 V_{k_i})
&\rightarrow& \Hom_{\Spin(4)}(\C,\bigotimes_{i=1}^4 V_{(j_i^-,j_i^+)}) \nn \\
\hat{\iota} &\mapsto& \phi(\hat{\iota}) := \iota,
\eea
embedding the vector space of $\SU(2)$ intertwiners into the vector space of $\Spin(4)$ intertwiners.

Explicitly, $\phi$ is constructed by using  the Clebsch-Gordan interwining maps $C^{j^- j^+}_{k} \colon V_k\rightarrow V_{j^-} \otimes V_{j^+}  $ injecting the $\SU(2)$
representation $V_k$ into the highest (resp. lowest) diagonal $\SU(2)$ subgroup factor $k = j^+ + j^-$ (resp. $k = j^+ - j^-$) of $V_{(j^-,j^+)} \cong V_{j^-} \otimes
V_{j^+}$ in the $\gamma < 1$ (resp. $\gamma > 1$) case. The labels $j^{\pm}$ and $k$ are related via the Immirzi parameter
by
\be
j^{\pm} = \frac{1}{2} |1 \pm \gamma| \, k.
\ee
These relations of course constrain the values of $k$ so that the $j^\pm$ are always half integers; specifically if $\gamma=p/q$ is written in lowest terms, then $k$ has to be a multiple of either $q/2$, or $q$ in some cases.

\begin{figure}
\begin{center}
\psfrag{a}{$ \int_{\Spin(4)} \, dG $}
\psfrag{b}{$\hat{\iota}^{k_1 k_2 k_3 k_4}$}
\psfrag{j1+}{$ j^+_{1}$}
\psfrag{j1-}{$ j^-_{1}$}
\psfrag{j2+}{$ j^+_{2}$}
\psfrag{j2-}{$ j^-_{2}$}
\psfrag{j3+}{$ j^+_{3}$}
\psfrag{j3-}{$ j^-_{3}$}
\psfrag{j4+}{$ j^+_{4}$}
\psfrag{j4-}{$ j^-_{4}$}
\psfrag{k1}{$ k_1$}
\psfrag{k2}{$ k_2$}
\psfrag{k3}{$ k_3$}
\psfrag{k4}{$ k_4$}
\psfrag{g+}{$ X^+$}
\psfrag{g-}{$ X^-$}
\psfrag{c1}{$ C^{j_1^+ j_1^-}_{k_1}$ }
\psfrag{c2}{$ C^{j_2^+ j_2^-}_{k_2}$ }
\psfrag{c3}{$ C^{j_3^+ j_3^-}_{k_3} $}
\psfrag{c4}{$ C^{j_4^+ j_4^-}_{k_4} $}
\includegraphics[scale=0.5,trim=0mm -10mm -40mm -10mm]{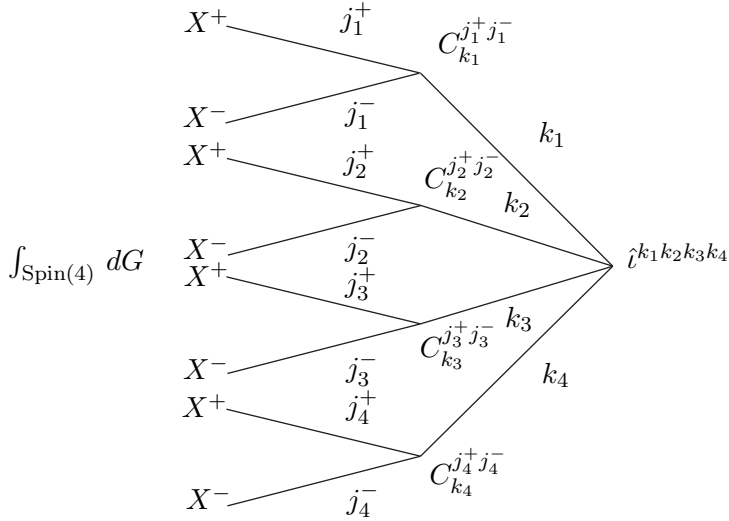}
\caption{The $\Spin(4)$ intertwiner $\iota$.}\label{figone}
\end{center}
\end{figure}

The $\Spin(4)$ intertwiner $\iota$ is then obtained as
follows
\be
\iota := \phi (\hat{\iota}) = \int_{\Spin(4)} \, dG \,\, (\pi_{j_i^-} \otimes \pi_{j_i^+})(G)
\circ \bigotimes_{i=1}^4 C^{j_i^- j_i^+}_{k_i}
\circ
\hat{\iota}^{k_1 k_2 k_3 k_4} \,\,
,
\ee
where the notation $G=(X^-, X^+)$ is used (see figure \ref{figone}). The group integration ensures that the resulting object is $\Spin(4)$-invariant, i.e., is an  element of
$\Hom_{\Spin(4)}(\C,\bigotimes_{i=1}^4 V_{(j_i^-,j_i^+)})$.

The dynamics of the EPRL model is encoded in the four-simplex, or vertex amplitude $f_4$ constructed by contracting the specified intertwining operators associated to each of the five tetrahedra $\Delta_3$ of the four-simplex $\Delta_4$. Labelling these tetrahedra by $a=1,...,5$, the ten triangles $\Delta_2$ of $\Delta_4$ are then indexed by the pair $ab$ of tetrahedra which intersect on the triangle. There are two $\SU(2)$ group elements
$(X_a^-,X_a^+)$ and one $\SU(2)$ intertwiner $\hat{\iota}_a$ for each tetrahedron, and three ($\gamma$-related)
$\SU(2)$ representations $k_{ab}$ and $(j_{ab}^-,j_{ab}^+)$ for each triangle. The intertwiner $\hat{\iota}_a$ lies in the space
$$\Hom_{\SU(2)}(\C,\bigotimes_{b:b\ne a} V_{k_{ab}}),$$
with the tensor product ordered by the numerical order of $b$, and $k_{ab}=k_{ba}$.

The amplitude $f_4\in\C$ is defined by forming a closed spin network diagram from this data. The five intertwiners (vertices) $\iota_{a}$ are tensored together and then the free ends are joined pairwise according to the combinatorics, i.e. the edge $a$ of vertex $b$ is joined to edge $b$ of vertex $a$. This is done using the standard `$\epsilon$ inner product' of irreducible representations of $\SU(2)$, denoted
$$\epsilon_k\colon V_k\otimes V_k\to\C.$$
This is defined by a choice of the two-dimensional antisymmetric tensor $\epsilon$ for $\SU(2)$ spin $1/2$, and extended to arbitrary spin by tensor products of $\epsilon$. This inner product is represented in the spin network diagram as a semicircular arc. To combine the $\Spin(4)$ intertwiners $\iota_{a}$, one regards each vertex as an $\SU(2)$ spin network and uses one $\epsilon$ inner product to connect the $j^+$ edges and a second $\epsilon$ inner product to connect the $j^-$ edges. The resulting closed diagram is evaluated using the binor calculus rules for all crossings. (Note: one could also use the rule that there is no sign for a crossing of a $+$ line with a $-$. This makes at most a difference of an overall sign to $f_4$.)

This yields a formula
\be
\label{amplitude}
f_4 = (-1)^\chi\int_{\Spin(4)^5} \prod_{a} d X^+_a d X^-_a
\left(
\bigotimes_{a<b}\mathcal{K}_{ab}
\right)
\circ
\left(
\bigotimes_a\hat{\iota}_a
\right)
\ee
where the propagators $\mathcal{K}_{ab}\colon V^{k_{ab}}\otimes V^{k_{ab}}\to\C $ are
defined by
$$
\mathcal{K}_{ab} =
\epsilon_{j^-_{ab}}\otimes \epsilon_{j^+_{ab}}\circ
\left(
\left(
 \pi_{j_{ab}^-}(X^-_{a}) \otimes \pi_{j_{ab}^+} (X^+_{a}) \,
\circ
C^{j_{ab}^- j_{ab}^+}_{k_{ab}} \,
 \right)\otimes
 \left(
 \pi_{j_{ab}^-}(X^-_{b}) \otimes \pi_{j_{ab}^+} (X^+_{b}) \,
\circ
C^{j_{ab}^- j_{ab}^+}_{k_{ab}} \,
\right)\right)
$$
and $(-1)^\chi$ is the sign defined by the diagrammatic calculus of spin networks.
Notice that as $f_4$ is linear in the $\hat{\iota}$ we can use unnormalized intertwiners and push the normalization into $f_3(\hat{\iota})^2 = \frac{1}{\hat{\iota} \, . \, \hat{\iota}}$, the asymptotic behaviour of which can then be analysed independently.

The aim of this paper is to study the large spin behaviour $k_{ab} \rightarrow \infty$ of the four-simplex amplitude $f_4$. Our strategy is to approximate the integral formula of the amplitude using (extended) stationary phase. At this level, the problem is not well posed because the scaling of the $\SU(2)$ intertwiners is not yet defined. Solving this problem naturally leads to a reformulation of the integral formula to an exponential form which is particularly well suited to asymptotics.

As a final remark, although we will not use this in the following, using the $\SU(2)$-invariance of the Clebsch-Gordan maps, one can set one of the two group
arguments of the propagator, say the left handed part, to the identity. In this case, the amplitude
\eqref{amplitude} becomes the Feynman evaluation associated to a tensor field theory over $S^3 \cong \SU(2)$. The
`matter fields' are identified as sections of the vector bundle $E = P \times_k V_k$ associated to the trivial
principal bundle $P = \Spin(4) \cong S^3 \times \SU(2)$ with base manifold $S^3$ and structure group $\SU(2)$ via
the representation $k$. The internal indices of the propagators of the matter fields are contracted at the
vertices of the diagram using the $\SU(2)$ intertwiners $\hat{\iota}$, and the amplitude \eqref{amplitude}, with
$X^- = 1 \!\! 1$, is the Feynman evaluation associated to the complete graph with five vertices.

\section{Posing the asymptotic problem}\label{posing}

The asymptotic large spin limit can be investigated by scaling each spin simultaneously, replacing
$$k_{ab} \rightarrow \lambda k_{ab},$$
 then fixing numbers $k_{ab}$, and taking the limit $\lambda \to \infty$ through the natural numbers. However to specify the boundary data correctly for each $\lambda$, one needs to `scale' the corresponding intertwiners. To do this, the intertwiners are specified in a particular way which gives a sequence of intertwiners in the corresponding spaces. The intertwiners are specified with coherent states, following the construction in \cite{livine-2007-76}.

\subsection{Coherent states}
Let $\mathbf n \in S^2$ be a unit 3-vector. Then a coherent state  $\alpha \in V_k$ in direction $\mathbf n$ is a unit vector satisfying
$$ (\mathbf L.\mathbf n) \, \alpha = i k \, \alpha,$$
with $\mathbf L$ the standard anti-hermitian rotation generators in the $k$ representation and the dot `.' denoting the 3d (Euclidean) scalar product. The coherent state has the maximal spin projection along the $\mathbf n$ axis.

One way of thinking of coherent states is to pick an arbitrary coherent state $\Gamma(\mathbf n)$ for each unit vector $\mathbf n$. Then any other coherent state is a phase factor $e^{i\theta}$ times one of these standard ones, $ \alpha=e^{i\theta}\Gamma(\mathbf n)$. This information is displayed in bra-ket notation as
$$\alpha=|k,\mathbf n,\theta\rangle.$$
In fact the choice of section $\Gamma$ does not play any role in the following.

The notation is shortened in the following ways. If the parameter $k$ is omitted, then the fundamental representation is used; thus $|\frac12,\mathbf n,\theta\rangle=|\mathbf n,\theta\rangle$. Further, the parameter $\theta$ will be omitted later when the phase is obvious from the context.

Embedding $V_k$ in $\otimes^{2k} V_{1/2}$ shows that
$$\otimes^{2k}|\mathbf n,\theta \rangle$$
is a coherent state for $\mathbf n$ and $V_k$. Therefore the Hermitian inner product of coherent states obeys
\be
| \la k, \mathbf n_1,\theta_1 | k , \mathbf n_2, \theta_2 \ra |^2 = \left( \frac{1}{2} (1 + \mathbf n_1 . \mathbf n_2) \right)^{2k},
\ee
where $| \, . \, |^2$ denotes modulus square.

\subsection{Coherent tetrahedra}

To construct a coherent intertwiner for a given tetrahedron, the idea is to associate a coherent state to each one
of its triangles. The geometrical picture is that the coherent state $|k, \mathbf n, \theta \ra$ then carries the interpretation
of the normal of length $k$ and direction $\mathbf n$ to the associated triangle, plus a phase factor.

Furthermore, we want to describe tetrahedra with three-dimensional rotational symmetry and the coherent intertwiners are thus constructed by integrating over all spatial directions the tensor product of four coherent states
\be \hat{\iota}^{k_1k_2k_3k_4}(\mathbf n_1,\mathbf n_2,\mathbf n_3,\mathbf n_4) = \int_{\SU(2)} dh \,\bigotimes_{i=1}^4 h | k_i, \mathbf n_i, \theta_i \ra . \ee
These intertwiners were introduced by Livine and Speziale \cite{livine-2007-76}, who gave an asymptotic formula for their normalisation.

According to the `quantization commutes with reduction' theorem of Guillemin and Sternberg \cite{GS}, the space of intertwiners is spanned by the $\hat \iota$ determined by vectors satisfying the {\em closure constraint}
\be k_1 \mathbf n_1+ k_2 \mathbf n_2+k_3\mathbf n_3+k_4\mathbf n_4=0\ee
Therefore in this paper the coherent tetrahedron states are always taken to satisfy this condition. The condition also implies there is a tetrahedron $t\in\R^3$ in Euclidean space, with the standard metric, which has these four vectors as the outward face normals and triangle areas equal to $k_i$ \cite{Baez:1999tk}. Throughout this paper attention is restricted to the cases where the tetrahedra are non-degenerate. This means that the analysis excludes a few extremal cases.

The tetrahedron $t$ is uniquely specified by the four vectors $\mathbf n_1,\mathbf n_2,\mathbf n_3,\mathbf n_4$, up to parallel translation. Therefore the tetrahedron has a definite orientation.
The coherent state $\hat{\iota}$ is averaged by the action of $\SU(2)$ on the coherent states, which amounts to an action of $\SO(3)$ on the tetrahedron $t$, i.e. rigid rotations which preserve the orientation. The geometry of the tetrahedron which is invariant under these rotations is a metric and an orientation. The coherent state is a quantum version of the geometry of this tetrahedron.

\subsection{Coherent boundary 3D manifolds}

\subsubsection{Boundary data}
For the asymptotic formula in this paper, the boundary manifold is just the boundary of the 4-simplex, i.e. a three-sphere triangulated with five tetrahedra. However the considerations of this section apply to more general cases. So consider $\Sigma$ to be a triangulated closed 3-manifold. To specify a coherent state for the whole of $\Sigma$, the information required is: i) a choice of spin $k$ for each triangle, ii) for each tetrahedron $\tau$, a choice of coherent state $|k,\mathbf n,\theta\rangle$ for each of its four triangles. Clearly the $k$ used in ii) are the ones picked in i). Such a choice $\mathcal B=\{k,\mathbf n\}$ for the whole of $\Sigma$ is called {\em boundary data} for the manifold $\Sigma$.

Then the boundary state specified by this data is
$$\psi(k,\mathbf n)=\bigotimes_{\text{tetrahedra}}\hat{\iota}^{k_1k_2k_3k_4}(\mathbf n_1,\mathbf n_2,\mathbf n_3,\mathbf n_4).$$
However the phase of this state vector is arbitrary, as the phase of each coherent state has not been specified yet. In the following, it is shown that for an important subset of boundary data, there is a canonical choice of phase.

\subsubsection{Regge-like boundary data}\label{Regge-like}

For each tetrahedron $\tau_a \subset \Sigma$ let $\phi_a: \tau_a \rightarrow \mathbb{R}^3$ be the affine linear map such that $\phi_a(\tau_a) = t_a$ is the geometric tetrahedron defined above, i.e., such that $n_{ab}$ is the outward normal of $t_a$ in the direction of the neighbouring tetrahedron $b$.

Boundary data will be called {\em Regge-like} if
\begin{itemize}
\item
There is a metric $g_{\Sigma}$ on $\Sigma$ (as in Regge calculus) such that the metric restricted to the tetrahedron $\tau_a$ is the pull-back of the standard metric $\delta$ on $\mathbb{R}^3$ with $\phi_a$, i.e. $g_{\Sigma}|_{\tau_a} = \phi_a^* \delta$.
\item
There is an orientation $\mathrm{Or}$ of $\Sigma$ such that the orientation
$\mathrm{Or}_{\tau_a}$ of each $\tau_a$ is the pull-back of the standard orientation on $\mathbb{R}^3$ with $\phi_a$.
\end{itemize}

Of course these conditions can only be satisfied if $\Sigma$ is an orientable manifold.

\subsubsection{Regge states}\label{Regge states}
For Regge-like boundary data there is a canonical choice of phase for the state, which will now be described.

The construction needs the standard antilinear structure map for representations of $\SU(2)$, $J\colon V_k\to
V_k.$ This is defined by
$$\epsilon_k(\alpha,\alpha')= \la J \, \alpha | \alpha' \ra,$$
the left-hand side being the epsilon-inner product and the right hand side the Hermitian inner product. It obeys
$Jg=gJ$ for all $g\in\SU(2)$, $J^2=(-1)^{2k}$ and $\langle
J\alpha|J\alpha'\rangle=\overline{\langle\alpha|\alpha'\rangle}$. Since
$$ J(i \mathbf n\cdot\mathbf L)= - (i \mathbf n\cdot \mathbf L) J,$$
the map $J$ takes a coherent state for $\mathbf n$ to a coherent state for $-\mathbf n$.

 Recall that the $\SU(2)$ coherent state $| k , \mathbf n , \theta \rangle$ represents a vector $\mathbf n$ of fixed length in $\R^3$, interpreted as describing the normal vector to a triangle lying in the plane $\mathbf n^{\bot}$ orthogonal to $\mathbf n$ in $\R^3$. This triangle is only defined up to $\U(1)$ rotations in the corresponding plane (this is the phase information), and the coherent state $| k , \mathbf n , \theta + \vphi \rangle$ describes the same triangle but rotated by an angle $2 \vphi$ in the $\mathbf n^{\bot}$ plane.

Let $\tau_a$ and $\tau_b$ denote two tetrahedra in $\Sigma$ which share a common triangle, $\Delta_{ab}=\tau_a
\cap \tau_b$.
The important point is that for Regge-like boundary data the geometries of the common triangle agree. This means
that there is a unique element $\hat g_{ab}\in \mathrm{O}(3)$ which, together with a translation, maps the image
of this triangle $\phi_a(\Delta_{ab})$ in $t_a$ congruently to the corresponding triangle $\phi_b(\Delta_{ab})$ in
$t_b$ and takes one outward-pointing normal to minus the other one,
 \bea\label{g_ab} \hat g_{ab} \circ \phi_a( \Delta_{ab} )&=& \phi_b( \Delta_{ab} )\nn\\
 \hat g_{ab} \mathbf n_{ab}&=& -\mathbf n_{ba}.\eea

In fact, $\hat g_{ab}$ is in $\SO(3)$, due to the following argument. Consider any two linearly independent
vectors $\mathbf v, \mathbf w$ in $\Delta_{ab}$. The element $ \hat g_{ab}$ is defined to be the linear
transformation that maps $\phi_a(\mathbf v)$ to $\phi_b(\mathbf v)$, $\phi_a(\mathbf w)$ to $\phi_b(\mathbf w)$
and $\mathbf n_{ab}$ to $-\mathbf n_{ba}$. However, due to the fact that $\phi_a$ and $\phi_b$ are
orientation-preserving, these vectors form frames that have the same orientation in $\mathbb{R}^3$. Thus $\hat
g_{ab}\in\SO(3)$.

These group elements can actually be lifted to $g_{ab}\in \SU(2)$, using a choice of spin structure for the manifold $\Sigma$. Given a spin structure, the Levi-Civita connection for the metric $g_\Sigma$ lifts to a spin connection on the spin bundle. As the manifold is not smooth everywhere this is a slight extension of the usual notion, but is determined in a fairly straightforward way by smoothing the conical singularities on the 1-skeleton of $\Sigma$ (the edges and vertices) and applying the usual definition to the smoothing. In fact the holonomy of a loop around a conical singularity is not just a rotation matrix, but in fact is a rotation angle (which deforms continuously to zero when the loop is shrunk to a point in a smoothing of the cone). In particular this determines uniquely the holonomy in the spin group around a conical singularity.

The Levi-Civita connection determines a parallel transport operator from the tangent space of one tetrahedron $\tau_a$ to another, $\tau_b$
$$\omega_{ab}: T\tau_a\to T\tau_b.$$
The link with the matrix $\hat g_{ab}$ is to regard it as the parallel transport operator in bases
provided by the tangent space frames $\phi_a$ and $\phi_b$ for each tetrahedron, i.e.,
$$ \hat g_{ab}       =       \phi_b \omega_{ab}\phi_a^{-1},$$
disregarding the translation part of these maps.
Concretely, this can be regarded as the parallel transport along a dual edge, with the frames $\phi_a$ and $\phi_b$
at the corresponding dual vertices.

Given a choice of spin structure for $\Sigma$, there are two spin
frames for each tetrahedron that cover the given tangent space frames. Choosing one of these for each tetrahedron
then defines the $g_{ab}\in \SU(2)$ as the parallel transport operators for the spin connection using these spin
frames as bases.

The choice of phase for the boundary state $\psi$ is given by picking the phase of $\alpha_{ab}=|k_{ab},\mathbf
n_{ab},\theta_{ab}\rangle$ for $\tau_a$ to be arbitrary, and then fixing the phase of the state for the
corresponding triangle in $\tau_b$ to be
 $$\alpha_{ba}= g_{ab}J\alpha_{ab}.$$

In the definition of $\psi$, both of these states appear; $\psi$ is an integral over a product of terms
$$\alpha_{ab} \otimes g_{ab}J\alpha_{ab},$$
one for each triangle. Since $J$ is antilinear, a change of phase of $\alpha$ cancels between these two factors of
the tensor product. This provides a canonical choice of the phase for the boundary state $\psi$. Regge-like
boundary data together with this choice of phase of the state $\psi$ is called a {\em Regge state}.

Although the $g_{ab}\in \SU(2)$ depend on the choice of a spin frame for each tetrahedron, changing the choice of
spin frame does not affect the Regge state. This is because changing the spin frame for tetrahedron $a$ to the
other possible choice results in the simultaneous change $g_{ab}\to -g_{ab}$ for all tetrahedra $b$ neighbouring
$a$. Since the sum of the spins on the face of tetrahedron $a$ is an integer, the action of $\otimes_b (-g_{ab})$
on the tensor product of the coherent states is the same as the action of  $\otimes_b g_{ab}$. Thus the Regge
state is the same for either of the two possible choices of spin frame at any given tetrahedron. It depends only
on the chosen spin structure. Of course, for the case of a 4-simplex, the boundary $\Sigma=S^3$ has only one spin
structure and this dependence is of no consequence.

It is worth noting that, in this context of general boundaries $\Sigma$, Regge states will behave well under
gluing. In other words, gluing simplexes together with the canonical choice of phase will produce a formula for
the partition function for a general manifold with the canonical choice of phase on the boundary.
\begin{figure}
\begin{center}
\psfrag{a}{$\ \  < k_{ab} ,  -\mathbf n_{ab} |$}
\psfrag{b}{$|k_{ab} ,  \mathbf n_{ba}>$}
\psfrag{c}{$(X^+_a)^{-1} $ }
\psfrag{d}{$X^+_b $ }
\psfrag{e}{$(X^-_a)^{-1} $ }
\psfrag{f}{$X^-_b $ }
\psfrag{g}{$k_{ab}$}
\psfrag{h}{$k_{ab}$}
\psfrag{i}{$j^+_{ab}$}
\psfrag{j}{$j^-_{ab}$ }
\psfrag{G}{$(h_a)^{-1}$}
\psfrag{H}{$h_b$}
\includegraphics[scale=0.5,trim=0mm -25mm -50mm -25mm]{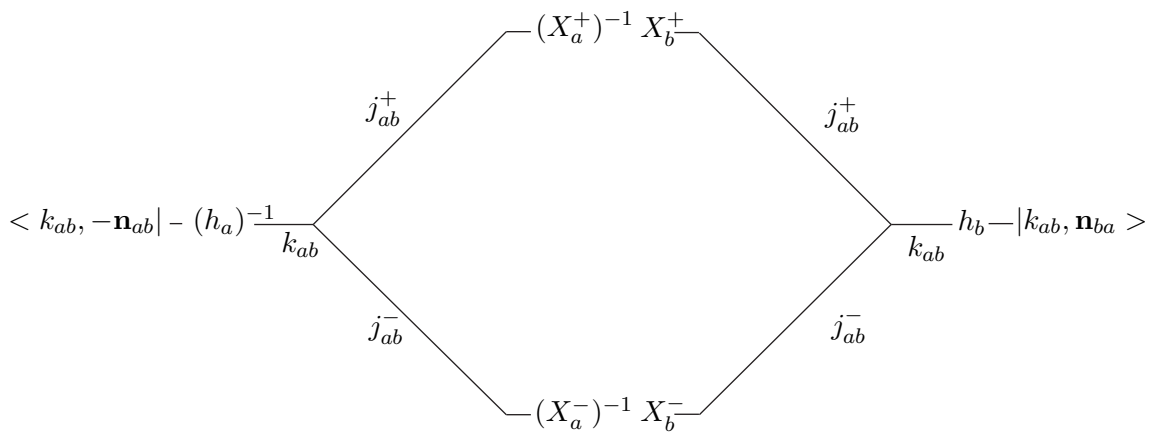}
\caption{The propagator $\mathcal{P}_{ab}$ for a single edge.}
\label{propagator diagram}
\end{center}
\end{figure}

From now on the notation for coherent state vectors will be shortened to $|k,\mathbf n\rangle$, omitting the phase
parameter. The notation $|k, -\mathbf n\rangle$ means $J|k,\mathbf n\rangle$.

\subsection{Exponentiated expression for the four-simplex amplitude}\label{Action Symmetries}

The general considerations of the previous sections are now applied, for the rest of the paper, to the case of a 4-simplex $\sigma$. The boundary data is specified on the simplicial 3-manifold $\Sigma=\partial\sigma$. Using the coherent states framework,
the four-simplex amplitude is
\be
\label{amplitude1}
f_4 = (-1)^{\chi'} \int_{\Spin(4)^5} \prod_{a} dG_a \, \int_{\SU(2)^5} \prod_a dh_a \prod_{a<b} \mathcal{P}_{ab} ,
\ee
where the coherent propagator $\mathcal{P}_{ab}$ is now given by
\be
\mathcal{P}_{ab} = \langle k_{ab}, -\mathbf n_{ab} |
 \pi_{k_{ab}}(h_a^{-1})
C_{ j_{ab}^- j_{ab}^+}^{k_{ab}} \,
\pi_{j_{ab}^-}(X^-_{ab}) \, \pi_{j_{ab}^+} (X^+_{ab}) \,
 C_{ k_{ab}}^{j_{ab}^- j_{ab}^+}
\pi_{k_{ab}}(h_b)
 | k_{ba}, \mathbf n_{ba} \rangle,
\ee
using the notation $X^{\pm}_{ab} := (X^{\pm}_a)^{-1} X^{\pm}_b$, and $C_{ j^- j^+}^{k}\colon V_k\to V_{j_-}\otimes V_{j_+}$ the reflected spin network, as shown in figure \ref{propagator diagram}.
This is proved starting from \eqref{amplitude}, rotating some of the Clebsch-Gordan coefficients using the epsilon inner products, then converting these inner products to the Hermitian inner product, and also flipping the order of $+$ and $-$ lines, obtaining further factors of $-1$ which are absorbed into the definition of $\chi'$.

The next step is to obtain an exponentiated version of the amplitude as a means to enter the framework of (extended) stationary phase. This is realized through a reformulation of the propagators.

The first remark in order is that the integration over  $\SU(2)$ in equation \eqref{amplitude1} at each vertex can be
absorbed into the $\Spin(4)$ integration because of the invariance of the Clebsh-Gordan maps. Then, the idea is to
use the exponentiating property of the coherent states
\be |k, \mathbf n \rangle =   | \mathbf n \rangle^{\otimes 2k}, \ee
 to reduce the propagator
to a product of two propagators in the fundamental representation of $\SU(2)$ to the power $2 j_{\pm}$
respectively.

For the next step, we need to treat the $\gamma < 1$ and $\gamma > 1$ cases separately.

\paragraph{$\gamma < 1$ case.}

\begin{figure}[ht]
\begin{center}
\psfrag{a}{$ k_{ab}= j^+_{ab} + j^-_{ab}$}
\psfrag{d}{$\vdots$ }
\psfrag{e}{$\vdots $ }
\psfrag{f}{$\vdots$ }
\psfrag{b}{$j^+_{ab}$}
\psfrag{c}{$j^-_{ab}$ }
\includegraphics[scale=0.4,trim=0mm -10mm 0mm -10mm]{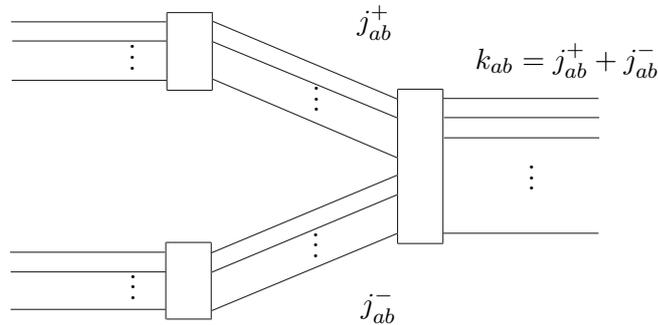}
\caption{The three-valent intertwiner $C^{j_{ab}^- j_{ab}^+}_{k_{ab}}$ for the case $\gamma <1$ showing how the projection to the highest spin subspace $k_{ab}$ makes two of the symmetrizers  redundant.}
\label{clebsch gordan fig}
\end{center}
\end{figure}

For $\gamma < 1$, the Clebsch-Gordan coefficient  $C^{j_{ab}^- j_{ab}^+}_{k_{ab}}$ injects into the highest
spin subspace $k_{ab}=j^+_{ab}+j^-_{ab}$.  Considered as a spin network, is easy to see that the symmetrizers on
the $j^+_{ab}$ and $j^-_{ab}$ edges can be absorbed into the symmetrizer on the $k_{ab}$ edge because of the
stacking property of symmetrizers, see figure \ref{clebsch gordan fig}.
The remaining symmetrizer now acts on the coherent  states $ | k_{ab}, \mathbf n_{ab} \rangle$ but since these are defined
as the symmetrized tensor product of spin half coherent states $ | \mathbf n_{ab} \rangle$ this final symmetrization can
also be ignored. We can now use the exponentiating property of the coherent states to further split the propagator
into a product of terms in the fundamental representation. We obtain the following expression for the propagator
\be
\mathcal{P}_{ab}^{\gamma < 1} = \la -\mathbf n_{ab}| X^-_{ab} |  \mathbf n_{ba} \ra^{2j_{ab}^-} \; \la -\mathbf n_{ab}| X^+_{ab} |
\mathbf n_{ba} \ra^{2j_{ab}^+}.
\ee

The four-simplex amplitude can thus be re-expressed as
\be \label{amplitude2}
f_4 = (-1)^{\chi '} \int_{\Spin(4)^5} \prod_{a} dG_a \; e^{S_{\gamma < 1}} ,
\ee
with the action given by
\be
\label{action}
S_{\gamma < 1} = \sum_{a < b} \, 2j_{ab}^- \, \ln \, \la -\mathbf n_{ab}| X^-_{ab} |   \mathbf n_{ba} \ra + 2j_{ab}^+ \, \ln \,
\la -\mathbf n_{ab}| X^+_{ab} |   \mathbf n_{ba} \ra.
\ee
Notice that this action is in general complex.  The logarithm of a complex number is only defined up to a multiple of $2\pi i$, we can safely neglect this factor as it will not affect the stationary points and when it appears in the action it is exponentiated.

\paragraph{$\gamma > 1$ case.}

\begin{figure}[ht]
\begin{center}
\psfrag{a}{$ k_{ab}= j^+_{ab} - j^-_{ab}$}
\psfrag{d}{$\vdots$ }
\psfrag{e}{$\vdots $ }
\psfrag{f}{$\vdots$ }
\psfrag{h}{$\ldots$ }
\psfrag{b}{$j^+_{ab}$}
\psfrag{c}{$j^-_{ab}$ }
\includegraphics[scale=0.4,trim=0mm -10mm 0mm -10mm]{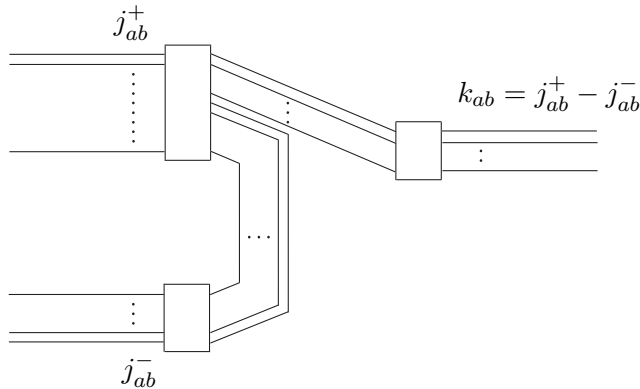}
\caption{The three-valent intertwiner $C^{j_{ab}^- j_{ab}^+}_{k_{ab}}$ for the case $\gamma >1$ projecting to the lowest weight state $k_{ab} = j^+_{ab} - j^+_{ab}$.}
\label{clebsch gordan fig gamma greater than one}
\end{center}
\end{figure}

The $\gamma > 1$ case works analogously but is more  complicated as the Clebsch-Gordan coefficient $C^{j_{ab}^- j_{ab}^+}_{k_{ab}}$ now injects into the lowest state $k_{ab}=j^+_{ab}-j^-_{ab}$, see figure \ref{clebsch gordan fig gamma greater than one}.

The symmetrizers on the $k_{ab}$ and $j^-_{ab}$ edges  can be absorbed into the symmetrizer on the $j^+_{ab}$
edge.  We remove the remaining symmetrizer by using the resolution of the identity in terms of $\SU(2)$ coherent
states in the $j^+_{ab}$ representation \cite{perelomov}
\be
1_{j} = d_j\int_{\SU(2)/\U(1)} d{\mathbf{n}} |j,\mathbf n \rangle \langle
j,\mathbf n|
\ee
where an arbitrary choice of phase is made for each coherent state, $d_j = 2j+1$, and the integral measure is normalised to $1$.

With these insertions, all of the symmetrizers can be absorbed into coherent states
and the propagator can be divided into terms in the fundamental representation as before
\bea
\label{gamma > 1 propagator}
\mathcal{P}_{ab}^{\gamma > 1} &=& (-1)^{2j_{ab}^-} d_{j^+_{ab}}^2
    \int d\mathbf{m}_{ab} \; \int d\mathbf{m}_{ba} \;
    \la -\mathbf n_{ab}|  \mathbf m_{ab} \ra^{2k_{ab}}
    \la \mathbf m_{ab}| X^+_{ab} | \mathbf m_{ba} \ra^{2j_{ab}^+} \;
     \nn \\ && \times
    \overline{\la \mathbf m_{ab}| X^-_{ab} |\mathbf m_{ba} \ra}^{2j_{ab}^-} \;
    \la \mathbf m_{ba}|  \mathbf n_{ba} \ra^{2k_{ab}}.
\eea
The action is given by
\bea
\label{gamma > 1 action}
S_{\gamma > 1}
&=& \sum_{a < b} \,
  2j_{ab}^+ \ln \, \la\mathbf  m_{ab}| X^+_{ab} | \mathbf m_{ba} \ra
+ 2j_{ab}^- \ln \, \overline{\la \mathbf m_{ab}| X^-_{ab} | \mathbf m_{ba} \ra}
\nn \\
& &
+ 2k_{ab} \ln \,  \la -\mathbf n_{ab}| \mathbf m_{ab} \ra
+ 2k_{ab} \ln \,  \la \mathbf m_{ba}| \mathbf  n_{ba} \ra
\eea
Note that we could just have easily decomposed the propagator using only the $\mathbf{m}_{ab}$, for $a<b$.  However, using the second resolution of the identity gives a formula that treats the $\pm$ sectors in a more symmetric way.

The $j^-$ term is easiest to understand in terms of the graphic notation. We first flip over the left bend in the
diagram. This causes the factor of $(-1)^{2j^-}$ in equation \eqref{gamma > 1 propagator}. Then we can straighten out the ``$S$" by using
standard graphical calculus, which leads to a transposition of the group element. Written as equations in the
fundamental representation this reads $\epsilon_{ji} X^j_k \epsilon^{kl} = - \epsilon_{ij} X^j_k \epsilon^{kl} = -
(X^{-1})^l_i$, using the convention $\epsilon_{ij}\epsilon^{jk}=\delta^k_i$. In terms of the standard inner product this part of the diagram then leads to the matrix element
$$-\la\mathbf m_{ba}| (X^-_{ab})^{-1} | \mathbf m_{ab}
\ra = -\overline{\la \mathbf m_{ab}| X^-_{ab} | \mathbf m_{ba} \ra}.$$

Thus the amplitude for $\gamma > 1$ is given by
\be
\label{amplitude3}
f_4 = (-1)^{\chi'} \prod_{a<b} (-1)^{ 2j_{ab}^-}\; d_{j^+_{ab}}^2 \int_{\Spin(4)^5} \prod_{c} dG_c \; \int d\mathbf{m}_{ab} d\mathbf{m}_{ba} \;    e^{ S_{\gamma > 1}}.
\ee

\subsubsection{Symmetries of the action.}\label{Symmetries of the action}

The actions \eqref{action} and \eqref{gamma > 1 action} admit two types of symmetry which will be important later.
\begin{itemize}
  \item {\em Continuous.} A global $\Spin(4)$ transformation characterised by an element $(Y^-,Y^+)$ in $\Spin(4)$
acting on each $X_a^\pm$ by $X_a^\pm \rightarrow Y^\pm X_a^\pm$. This determines a rigid motion of the whole 4-simplex.
  \item {\em Discrete.} At each vertex $a$ the transformation $X^+_a \rightarrow -
X^+_a$ leaves a factor $(-1)^{\sum_{b ,b \neq a} 2j^+_{ab} }$.  Since
the $j^+_{ab}$ sum to an integer at each tetrahedron $a$, this factor
will equal one. This symmetry is a consequence of the fact that $\SU(2)$ is the double cover of $\SO(3)$.
      There is a similar symmetry for the minus sector, $X^-_a\to -X^-_a$.
\end{itemize}

\section{Asymptotic formula}\label{asymptotic}
To study the semi-classical approximation of the theory we consider the large spin limit, and study the asymptotics. We start by scaling all ten spins by a constant parameter $k_{ab} \rightarrow \lambda k_{ab}$. Then the scaled boundary state used is
$$\psi_\lambda=\psi(\mathbf n_{ab},\lambda k_{ab}).$$
If the boundary data is Regge-like then the Regge state is used here. Otherwise the phase of the state is undetermined.

The asymptotic behaviour of the 4-simplex amplitude $f_4$ depends on the geometric character of the boundary data.
Recall that it is assumed throughout that for each tetrahedron $\tau_a$, the four vectors $\mathbf n_{ab}$, $b\ne a$, are the outward normals to the faces of a non-degenerate tetrahedron $t_a$ in three-dimensional Euclidean space.

There are two important types of boundary data for a 4-simplex. The first is for Regge-like boundary data where the boundary metric is the boundary of a non-degenerate four-dimensional Euclidean geometry for the 4-simplex. In other words, the $t_a$ geometries fit together in four-dimensional space.

The second type which is important is called a vector geometry.
 The boundary data (not necessarily Regge-like) is said to be {\em vector geometry} if there is unit vector ${\bf v}_{ab} \in\R^3$ assigned to the $ab$-th triangle satisfying
$$\sum_{b: b\ne a} k_{ab}{\bf v}_{ab}=0$$
$${\bf v}_{ab} = - {\bf v}_{ba}$$
and for each tetrahedron, the $\mathbf n_{ab}$ are congruent to the corresponding ${\bf v}_{ab}$ (i.e. $\mathbf n_{ab}=h_a {\bf v}_{ab}$ for $h_a\in\SO(3)$).

Examples of vector geometries are determined by a non-degenerate 4-simplex. This gives two distinct vector geometries, given by the self-dual and anti-self-dual parts of the triangle bivectors.
Another example of a vector geometry is given by a linear map of a 4-simplex into $\R^3$.

In the case of a non-degenerate 4-simplex, the geometry of the simplex is uniquely defined by the geometry of the boundary, and hence by the boundary data. For this result in convex rigid geometry see for example Theorem 4.10 in \cite{rigidity}. Therefore there are well-defined dihedral angles $0<\Theta_{ab}<\pi$ for each triangle. These are the angles between the outward normals to the tetrahedra embedded in $\R^4$. These angles are used in the asymptotic formula.

A formula is asymptotic if the error term is bounded by a constant times one more power of $\lambda^{-1}$ than that stated in the asymptotic formula.

\begin{theo} {\bf (Asymptotic formula)}
\label{big theorem new}
Given a set $\mathcal{B}=\{ \mathbf n_{ab} , k_{ab} \}_{a \neq b}$ of boundary data, then in the limit
$\lambda \rightarrow \infty$
\begin{enumerate}
  \item  If  $\mathcal B$ is a non-degenerate 4-simplex geometry, and $\psi_\lambda$ is the associated Regge state, then
\begin{multline}
f_4(\psi_\lambda  )
\sim
(-1)^{\chi'}
\left(\frac{2 \pi}{\lambda}\right)^{12}
\left[
2 N^\gamma_{+-}\cos \left( \lambda\gamma\sum_{a<b}k_{ab}\Theta_{ab}\right)
\right. \\
+
N^\gamma_{++}\exp{\left(i \lambda\sum_{a<b}k_{ab}\Theta_{ab}\right)}
+
\left.
N^\gamma_{--}\exp{ \left( -i\lambda\sum_{a<b}k_{ab}\Theta_{ab}\right)}\right]
\end{multline}
The numbers $N^\gamma_{+-}$, $N^\gamma_{++}$, $N^\gamma_{--}$ are
independent of $\lambda$ and
are defined
below.

\item For a set of boundary data which is not a non-degenerate 4-geometry but forms a vector geometry
  the asymptotic formula is
\bea
|f_4(\psi_\lambda )|
&\sim&  \left(\frac{2 \pi}{\lambda}\right)^{12}
N
\eea
The number $N$ is independent of $\lambda$ and
is defined
below.

\item  For a set of boundary data that is neither a non-degenerate 4-geometry nor a vector geometry,
the amplitude is supressed and does not contribute for large
$\lambda$.
\be
f_4(\psi_\lambda ) = o(\lambda^{-M}) \; \; \; \forall M
\ee
\end{enumerate}
\end{theo}

\section{Proof of the asymptotic formula}\label{proof}

In this section, we provide a proof of the above theorem. We first review the notion of (extended) stationary phase,
showing how the asymptotic formula is dominated by the critical points. Next, we calculate these points and relate the non-degenerate critical points to geometric four-simplices in $\R^4$. Finally, we evaluate the integrand of the
integral formula on the non-degenerate critical points and show how to recover terms related to the Regge action for the corresponding four-simplex.

\subsection{Extended stationary phase}

To study the semi-classical approximation of the theory, we are interested in the limit of expression \eqref{amplitude2} or \eqref{amplitude3} when $\lambda \rightarrow \infty$. Our strategy is to use extended stationary phase methods, that is, stationary phase
generalized to (non purely imaginary) complex functions. The general approximation scheme is stated as follows \cite{Hormander}.

Let $D$ be a closed manifold of dimension $n$, and let $S$ and $a$ be smooth,  complex valued functions on $D$ such that the real part $\Reel S\le0$. Consider the function
 \be f(\lambda) = \int_D d x \, a(x) \, e^{\lambda
S(x)}. \ee
The Hessian of $S$ is the $n\times n$ matrix denoted $H$. The stationary points are assumed to be isolated and non-degenerate; $\det H \neq 0$.

In the extended stationary phase, the key role is played by {\em critical points}, that is, stationary points for which $\Reel S=0$. If $S$ has no critical points then
  for large
parameter $\lambda$ the function $f$ decreases faster that any power of $\lambda^{-1}$. In other words, for all $N
\geq 1$: \be \label{formula} f(\lambda) = o(\lambda^{-N}), \ee

If there are critical points, then each critical point contributes to the asymptotics of $f$ by a term of order
$\lambda^{-n/2}$. For large $\lambda$ the asymptotic expansion of the integral yields for each critical point\be
\label{formula1}  a(x_0) \left(\frac{2 \pi}{\lambda}\right)^{n/2} \frac
1{\sqrt{ \det (-H)}} \, e^{\lambda S(x_0)} \left[1+ O(1/\lambda) \right]. \ee
At a critical point, the matrix $-H$ has a positive definite real part, and the square root of the determinant of this matrix is the unique square root which is continuous on matrices with positive definite real part, and positive on real ones.

If $S$ admits several isolated critical points with non-degenerate Hessian, we obtain a sum of
contributions of the form \eqref{formula1} from each of them. So, to compute the dominant terms in  the asymptotics of \eqref{amplitude3} for large spins, we need to find the stationary points of the action $S$ and restrict to those with zero real part.

\subsubsection{Stationary points}

We separate the variational problems of the $\gamma < 1$ and $\gamma > 1$ sectors.

\paragraph{$\gamma < 1$ case.}

The stationary points of $S_{\gamma < 1}$, determined  by varying the group variables $G=(X^-,X^+)$, are the same
as the stationary points of
\be
I_{\gamma < 1} := e^{S_{\gamma < 1}} = \prod_{a < b} \la - \mathbf n_{ab}| X^-_{ab} |
\mathbf n_{ba} \ra^{2j_{ab}^-} \; \la -\mathbf n_{ab}| X^+_{ab} |  \mathbf n_{ba} \ra^{2j_{ab}^+}.
\ee
The variation of an arbitrary
$\SU(2)$ group element $X$, and the associated variation of its inverse $X^{-1}$ are given by
\be
\label{variation of X}
\delta X = T \circ X \;\;\;\; \Rightarrow \;\;\;\; \delta X^{-1} = - X^{-1} \circ T,
\ee
where
$T = \frac{1}{2}i T^r \sigma_r$ is an arbitrary element in the Lie algebra $\su(2)$, and $\sigma_r$ are the Pauli matrices, with eigenvalues $\pm1$, satisfying
$$\sigma_r\sigma_s=\delta_{rs}1\!\! 1+i\epsilon_{rst}\sigma_t.$$
The variation of
$I$ leads\footnote{Here, we are supposing that $\la \mathbf n_{ab} | X_{ab}^{\pm}| - \mathbf n_{ba} \ra \neq 0$, for all
$a<b=1,...,5$. } to two
 complex $3d$ vector equations for each tetrahedron $a$:
\be
\label{stationary}
\delta I_{\gamma < 1} = 0 \;\;
\Leftrightarrow \;\; \forall a \in \{1,...,5\}, \;\;\;\;\;\;\;\;\;\;
\sum_{b\colon b \ne a} j_{ab}^{\pm} \,\, \mathbf{V}_{ab}^{\pm} = 0,
\ee
with
\be
\label{vector}
\mathbf{V}_{ab}^{\pm} = \frac{\la -\mathbf n_{ab} |(X_a^{\pm})^{-1} \, {\pmb\sigma} \, X_b^{\pm} |   \mathbf n_{ba} \ra}{\la
-\mathbf n_{ab} | (X_a^{\pm})^{-1} X_b^{\pm} |   \mathbf n_{ba} \ra},
\ee
where we have introduced a bold letter notation for
three-dimensional vectors: ${\mathbf V} \equiv (V_1,V_2,V_3)$ and ${\pmb \sigma} = (\sigma_1,\sigma_2,\sigma_3)$.
The minus signs arising from \eqref{variation of X} can be absorbed by noting that $\mathbf{V}_{ab}^{\pm} = -\mathbf{V}_{ba}^{\pm} $ This can be proved by using the fact that the formula is the ratio of two epsilon inner products, and that whilst $\epsilon(g\alpha,\beta)=\epsilon(\alpha,g^{-1}\beta)$ for a group element $g$, for a Lie algebra element $T$ one has $\epsilon(T\alpha,\beta)=-\epsilon(\alpha,T\beta)$.

As remarked above, each of these equations  is complex. To extract a real and an imaginary part will need a
bit more work. Firstly, we use the fact that the action of an element of $\SU(2)$ on a coherent state produces a
new coherent state (temporarily putting the phase information back in the notation)
$$|\mathbf  n_{ab}^{\pm}, \theta_{ab}^{\pm} \ra :=
X_{a}^{\pm}| \mathbf n_{ab}, \theta_{ab} \ra, $$
with the three-vectors
$$\mathbf n_{ab}^{\pm}=X_{a}^{\pm}\mathbf n_{ab}.$$
In this formula the $\SU(2)$ element acts on a 3-vector via the homomorphism to $\SO(3)$.  We will use this notation for the remainder of the paper. 
Hence, the complex three-vector \eqref{vector} can be re-written as
\be
\label{vector1}
\mathbf{V}_{ab}^{\pm} = \frac{\la -\mathbf n_{ab}^{\pm} |\, {\pmb\sigma} \,|   \mathbf n_{ba}^{\pm} \ra}{\la -\mathbf n_{ab}^{\pm} |  \mathbf  n_{ba}^{\pm} \ra} = \frac{\la -\mathbf n_{ab}^{\pm} |\, {\pmb\sigma} \, |  \mathbf  n_{ba}^{\pm} \ra \la  \mathbf  n_{ba}^{\pm} |  -\mathbf  n_{ab}^{\pm} \ra}{|\la -\mathbf n_{ab}^{\pm} |  \mathbf  n_{ba}^{\pm} \ra|^2}.
\ee
Next, we will need the following lemma.

\begin{lem}\label{spinor-vector}
For all couples of coherent states $(|\mathbf n_1 \ra, |\mathbf n_2 \ra)$  associated to the couple $({\mathbf  n}_1,{ \mathbf n}_2)$ of
elements of $S^2$, the following identity holds:
\be
\la \mathbf n_1 | {\pmb \sigma} | \mathbf n_2 \ra \, \la \mathbf n_2 | \mathbf n_1 \ra =
\frac{1}{2} ( \mathbf{n}_{1} + \mathbf{n}_{2} - i \mathbf{n}_1 \times \mathbf{n}_2),
\ee
where the symbol `$\times$' denotes the three-dimensional cross-product.
\end{lem}

{\em Proof.} The projector $P_\mathbf n$ on the coherent state $|\mathbf n \ra$, can be written in terms of Pauli matrices
$$
P_\mathbf n = | \mathbf n \ra \la \mathbf n | = \frac{1}{2} ( 1\!\! 1 + {\pmb\sigma} \cdot\mathbf n ).
$$
It obviously satisfies idempotency $P_\mathbf n \circ P_\mathbf n = P_\mathbf n$  and unit trace $\tr P_\mathbf n = 1$. Hence, the left hand side
of the above equation reads
\bea
\la \mathbf n_1 | {\pmb \sigma} | \mathbf n_2 \ra \, \la \mathbf n_2 | \mathbf n_1 \ra &=& \frac{1}{2} \left( \la\mathbf  n_1 | {\pmb \sigma} | \mathbf n_1 \ra + \mathbf{n}_{2} \la \mathbf n_1 | 1 \!\! 1 | \mathbf n_1 \ra + i \mathbf{n}_{2} \times \la \mathbf n_1 | {\pmb \sigma} | \mathbf n_1 \ra \right) \nn \\
&=& \frac{1}{2} ( \mathbf{n}_{1} + \mathbf{n}_{2} - i \mathbf{n}_1 \times \mathbf{n}_2) ,
\eea
where we have used $\la\mathbf n|{\pmb\sigma} | \mathbf n \ra = {\mathbf n} $ and $\la \mathbf n |\mathbf  n \ra =1$ in the last step. $\square$

Using the above lemma, the complex three-vector  \eqref{vector1} can be split into real and imaginary parts, and
the stationary equations \eqref{stationary} become the following twenty real vector equations
\be
\label{stationarity}
\sum_{b\colon b \neq a} j_{ab}^{\pm} \,\, \frac{ { \mathbf n}_{ab}^{\pm} - {\mathbf n}_{ba}^{\pm}}{1 - {\mathbf
n}_{ab}^{\pm} \, . \, {\mathbf  n}_{ba}^{\pm}} = 0, \;\;\;\; \mbox{and} \;\;\;\; \sum_{b\colon b \neq a} j_{ab}^{\pm} \,\,
\frac{ {\mathbf n}_{ab}^{\pm} \times {\mathbf  n}_{ba}^{\pm}}{1 - {\mathbf n}_{ab}^{\pm} \, . \, {\mathbf n}_{ba}^{\pm}} = 0,
\ee
for all $a=1,...,5$.

\paragraph{$\gamma > 1$ case.}

The variation  for $\gamma > 1$ proceeds in the same way but in this case the
stationary equations  are
\be
\label{gamma > 1 stationary pts} \sum_{b \ne a} j^{\pm}_{ab}
\mathbf{V}_{ab}^{\pm} = 0
\ee
with
\be
\label{vector gamma >1}
\mathbf{V}_{ab}^{\pm} =
 \frac{\la \mathbf m_{ab} |(X_a^\pm)^{-1} \, {\pmb\sigma} \, X_b^\pm | \mathbf m_{ba} \ra}
{\la \mathbf m_{ab}| (X_a^\pm)^{-1} X_b^\pm | \mathbf m_{ba} \ra},
\ee
plus variational equations for $\mathbf m$ which are vacuous at the critical points, discussed in the next section.

\subsubsection{Critical points}

As already remarked, the action \eqref{action}  is complex, and standard stationary phase does not apply. The key
point is to supplement the stationarity condition with a maximisation condition on the real part of the action.
The restriction of the stationary points to those maximising the real part of the action are called critical
points. They contribute to the asymptotic formula while the other points are exponentially damped.  The integrand of the four-simplex amplitude \eqref{amplitude1} is of the
schematic form
$$
I \sim e^{\lambda a} \, \, e^{\lambda i b},
$$
where $a \leq 0$ is the real part of the action  and $b$ is the imaginary part. When $\lambda \rightarrow \infty$,
the only significant contribution to the integral comes from the configurations where $e^a = 1$, i.e., when $a$
hits its maximal value $a=0$.

\paragraph{$\gamma < 1$ case.}

More precisely, the real part of the action \eqref{action} in the $\gamma < 1$ case is given by
\be \mbox{Re} \,
S_{\gamma < 1} = \sum_{a < b} \, j_{ab}^- \, \ln \, \frac{1}{2} ( 1 - {\mathbf  n}_{ab}^{-} \cdot {\mathbf  n}_{ba}^{-}) +
j_{ab}^+ \, \ln \, \frac{1}{2} (1 - {\mathbf  n}_{ab}^+ \cdot {\mathbf n}_{ba}^+),
\ee
where we have used the expression of the  inner product between coherent states and all phases have been absorbed in the imaginary part of the action.
The maximum $\mbox{Re} \, S = 0$ of this expression is obtained when all ${\mathbf n}_{ab}^{\pm}$ and ${\mathbf n}_{ba}^{\pm}$ are {\em anti-parallel}, i.e., when for all $a<b = 1,...,5$,
\be
\label{Re} X_a^{\pm } {\mathbf  n}_{ab}
= - X_b^{\pm } {\mathbf  n}_{ba}.
\ee
Thus, the dominating points in the integrand of the four-simplex amplitude
\eqref{amplitude1} are given by the points satisfying stationarity \eqref{stationarity} {\em and} maximising the
real part of the action \eqref{Re}. All other points are exponentially suppressed. Evaluated on such points, the
second stationarity equation becomes vacuous, and the critical points are those satisfying the ten vector
equations \eqref{Re}, plus the five vector equations

\be
\label{critical} \sum_{b\colon b \neq a} k_{ab} \,\, {\mathbf  n}_{ab} = 0
\ee
for all $a=1,...,5$. We will refer to these two equations as orientation and closure respectively, for reasons that will become clear shortly.

\paragraph{$\gamma > 1$ case.}

The critical points for the $\gamma > 1$ case are only slightly more complicated.  The only significant contribution in the asymptotic limit comes when the following are satisfied
\bea
\label{gamma > 1 crit pts}
   - {\mathbf n}_{ab}      &=& {\mathbf  m}_{ab}      \nn \\
    {\mathbf m}_{ba}      &=& {\mathbf n}_{ba}    \nn \\
    X_a^+ {\mathbf m}_{ab}&=& X_b^+ {\mathbf m}_{ba}   \nn \\
    X_a^- {\mathbf m}_{ab}&=& X_b^- {\mathbf m}_{ba}.
\eea
for $a<b$.
However, these conditions are equivalent  to the
set of equations \eqref{Re}. The action \eqref{gamma > 1 action} at the critical points, when expressed as a function of $\gamma$ and the boundary data, gives the same function as \eqref{action} but for the different range of $\gamma$.

The integral coming from the resolution of the identity is no longer relevant as we assume the measures $d\mathbf m_{ab},d\mathbf m_{ba}$
are normalised to one.  Using equations \eqref{gamma > 1 crit pts} in \eqref{gamma > 1 stationary pts} along with the fact that $k_{ab} =j_{ab}^+ - j_{ab}^- $, one can see that the stationary point conditions for  $\gamma > 1$ are equivalent to those for $\gamma < 1$.

\subsection{Geometrical interpretation}

In this section, we show how geometrical structures emerge from the solutions to the critical point equations.

\subsubsection{Bivectors} \label{bivectorsection}

Let $\Lambda^2(\R^4)$ be the space of bivectors. A pair of vectors $N,M\in\R^4$ determines a simple bivector
$N\wedge M$, and a general bivector is a sum of simple bivectors. A simple bivector $B$ can be considered as an antisymmetric tensor
$$N\wedge M=N\otimes M - M\otimes N,$$
and hence an element of the Lie algebra $\so(4)$, using the Euclidean metric. This gives the action on a vector $Q$ as

\be
\label{bivaction}
(N\wedge M)Q = (M\cdot Q)N - (N\cdot Q)M.
\ee
The norm $|B|$ of a simple bivector $B$ is defined so that

\be
\label{bivnorm}
 |N\wedge M|=|N||M|\sin\theta,
\ee
with $\theta$ the angle between the vectors. Consequently, in terms of the components $B^{IJ}$ (for $I,J=0,\ldots ,3$) of the antisymmetric tensor $B$ this is
$$ |B|^2=\frac12 B^{IJ}B_{IJ},$$
the indices being raised or lowered using the standard Euclidean metric $\delta$.

Let $*$ be the Hodge operator acting on $\Lambda^2(\R^4)$; $(* B)^{IJ} = \frac{1}{2} \epsilon^{IJ}_{\;\;KL}
B^{KL}$, where $\epsilon_{IJKL}$ is the totally antisymmetric four-dimensional tensor ($\epsilon^{0123} = 1$). With this definition, the Hodge operator satisfies $**=1$ and preserves the norm of a bivector.

In the following it will be useful to split the bivectors into two three-vectors.
Consider the eigenspace decomposition of the Hodge operator $\Lambda^2(\R^4) \cong \Lambda^2_-(\R^4) \oplus \Lambda^2_+(\R^4)$ decomposing any bivector $B$ into self-dual ${\bf b}^+$ and anti-self-dual ${\bf b}^-$ components, $* {\bf b}^{\pm} = \pm {\bf b}^{\pm}$. We will extensively use the vector space isomorphism $\Lambda^2_{\pm}(\R^4) \cong \R^3$ explicitly given by
\be
\label{selfdual} { b}^{\pm \, i} =  \pm B^{0i} + \frac{1}{2}
\epsilon^i_{\; jk} B^{jk},
\ee
with $i,j,k = 1,2,3$.
Using the standard Euclidean inner product on $\R^3$, this gives
$$|\mathbf b^+|^2+|\mathbf b^-|^2=2|B|^2.$$
This means that for a simple bivector $B$, $|\mathbf b^+|=|\mathbf b^-|=|B|$.

If $B$ and $C$ are bivectors, then they form a Lie algebra with $[B,C]=BC-CB$, using the action \eqref{bivaction} on vectors.
The above isomorphism can be extended to an isomorphism of Lie algebras
$$\so(4) \cong  \Lambda^2(\R^4 )  \cong    \R^3\oplus\R^3   \cong   \su(2)\oplus\su(2),$$
using the map
$$(\mathbf b^-,\mathbf b^+)\mapsto(\mathbf b^-\cdot \mathbf L\;,\mathbf b^+\cdot\mathbf L)$$
and the notation $\mathbf{b\cdot L}=b^iL_i$ with $L_i$ the generators of $\su(2)$. A calculation shows that this is an isomorphism providing $L_j=\frac i2\sigma_j$ in the fundamental representation, so that
$$[L_i,L_j]=-{\epsilon_{ij}}^k L_k.$$

\subsubsection{The geometrical 4-simplex}

A geometrical four-simplex $\sigma$ is determined by the position of its vertices in $\R^4$. It is
said to be non-degenerate if the five vertices do not lie in a hyperplane. In this case one also refers to the closed set formed by the convex hull of the vertices as the geometric 4-simplex; these can be glued together to form Regge manifolds.

Each oriented triangle $\Delta$ of a geometric 4-simplex $\sigma$ determines a bivector $B_{\Delta}(\sigma)$ by wedging together two of its edge vectors. However the constructions presented in this paper naturally label the triangles of a 4-simplex by the two tetrahedra which intersect at the triangle. Suppose $N_a(\sigma)$ is the outward unit normal vector to tetrahedron $a$. Then $N_a(\sigma) \wedge N_b(\sigma)$ defines a bivector which is in the plane orthogonal to the triangle where tetrahedra $a$ and $b$ intersect. Therefore $*(N_a(\sigma)\wedge N_b(\sigma))$ lies in the plane of the triangle; normalising it correctly then gives the definition of the bivector for the triangle labelled by the ordered pair $ab$. If $k_{ab}$ is the area of triangle $ab$, then this bivector is
\be \label{geometric bivectors}
B_{ab}(\sigma)= k_{ab} * \frac{N_a(\sigma)\wedge N_b(\sigma)}{|N_a(\sigma)\wedge N_b(\sigma)|}.\ee

This definition has the property that  $B_{ab}(\sigma)=-B_{ba}(\sigma)$. Also, the bivectors are unchanged if $\sigma$ is replaced by $G \sigma$, where $G$ is an inversion $x\to -x$ of $\R^4$, or a parallel translation. Apart from these symmetries, the bivectors determine the 4-simplex geometry uniquely.

To state this theorem precisely, a general definition is required.
An arbitrary set of ten bivectors $B_{ab}$ is said to be a {\em bivector geometry} if the following constraints are satisfied

\begin{itemize}
\item {\em Orientation.} A bivector registers the orientation of the triangle to which it is associated, i.e.,
$$
B_{ab} = - B_{ba}.
$$
\item {\em Closure.} Each tetrahedron building up the four-simplex must close,
$$
\forall a, \;\;\;\; \sum_{b\colon b \neq a} B_{ab} = 0.
$$
\item {\em Diagonal simplicity.} Each bivector must define a geometric plane, i.e., is a simple bivector
$$B_{ab}\wedge B_{ab}=0 \;\;\;\; \Leftrightarrow \;\;\;\; |\mathbf b^+_{ab}|=|\mathbf b^-_{ab}|.$$
\item{\em Cross simplicity.} Each couple of bivectors
belonging to the same tetrahedron must define a three-dimensional hyperplane spanned by the two planes
associated to each bivector:
$$ \forall a, \;\; \exists N_a \in \R^4, \;\; \mbox{such that} \;\; N_{a
I} B_{ab}^{IJ} = 0, \;\; \forall b \neq a.
$$
\item {\em Tetrahedron.} For three triangles meeting at a vertex $e$ of the $a$-th tetrahedron $(bcde)$,
$$
\tr \, \left(B_{ab}[B_{ac},B_{ad}]\right) \ne0,
$$
where we have used the isomorphism $\Lambda(\R^4) \cong \so(4)$.
\item {\em Non-degeneracy.} The assignment of bivectors is non-degenerate. This means that for six triangles
sharing a common vertex, the six bivectors are linearly independent.
\end{itemize}

The following theorem is proved in
\cite{barrett-1998-39}, and more explicit detail is given on the tetrahedron constraint in \cite{Baez:1999tk}. One difference is that the constraints have been adapted to label the triangles by the missing vertices, which means that the bivectors depend on the orientation of the 4-simplex.

The cross simplicity constraint stated here is as recently reformulated in \cite{Engle:2007qf}. Note that the cross-simplicity constraint of \cite{barrett-1998-39} requires only that two bivectors $B_{ab},B_{ac}$ respect $B_{ab}\wedge B_{ac}=0$. This
implies for simple bivectors sharing an index that they either belong to the same three-dimensional hyperplane or share a common direction. However in the case of the common direction the tetrahedron condition is not satisfied, as shown in \cite{barrett-1998-39}. Hence the formulation given here is equivalent.

\begin{theo} {\bf (Bivector geometry)}
\label{bivectors}
The bivectors $B_{ab}(\sigma)$ of a geometric 4-simplex $\sigma$ satisfy the bivector geometry constraints. Conversely, given a set of bivectors $B_{ab}$ on an abstract 4-simplex satisfying these constraints, there is a parameter $\mu=\pm1$ and geometric 4-simplex $\sigma$, unique up to translation and inversion, such that $B_{ab}(\sigma)=\mu B_{ab}$.
\end{theo}

We now show how the bivector geometry theorem relates to our critical point equations. We start by discussing the general idea. Consider an arbitrary bivector $B = ({\mathbf b}^-, {\mathbf b}^+)$.
This bivector satisfies the diagonal simplicity constraint if and only if $| \mathbf b^- | = | \mathbf b^+ |$. This is equivalent \cite{Livine:2007ya} to stating that there exists a $\Spin(4)$ rotation $G=(X^-,X^+)$ such that $\mathbf b^{\pm} = X^{\pm} {\mathbf v}$, with ${\bf v}$ a fixed, arbitrary vector in $\R^3$.
If $G=(X,X)$ belongs to the diagonal $\SU(2)$  subgroup, the left and right handed components of $B$ are equal as
vectors ${\bf b}^- = {\bf b}^+$ and it is easy to see that a such bivector satisfies $\pole_{I} B^{IJ} = 0$, with
$\pole=(1,0,0,0)$ the $S^3$ `north pole'. In other words, $B$ lives in a hyperplane orthogonal to $\pole$. In the general case, if $G$ does not belong to the diagonal $\SU(2)$ subgroup, one can show that the bivector $B=(X^-,X^+)({\bf v},{\bf v})$ satisfies the equation $N_I B^{IJ} = 0$, with $N = G \pole$, and thus lies in a hyperplane orthogonal to the rotated vector $N$. To summarise, a bivector $B$ satisfying the diagonal simplicity constraint can always be written $B=(X^- {\bf v}, \, X^+{\bf v})$ with the spin group element $(X^-,X^+)$ encoding the information on the normal defining the hyperplane in which the corresponding bivector lives. This idea will prove instrumental in the geometrical interpretation of the critical points. To start with, we introduce the following lemma.

\begin{lem}
Let $\sigma$ be a geometric four-simplex in $\R^4$ with the areas of its triangles being half-integers. Then there exists a set of boundary data $\mathcal{B}$, together with a set $X_a^{\pm}$ satisfying closure \eqref{critical} and orientation \eqref{Re}, associated to $\sigma$. These are detemined uniquely up to the action of $\SU(2)$ at each tetrahedron.
\end{lem}

{\em Proof.} Consider a geometrical four-simplex $\sigma$. It is determined by five tetrahedra, each spanning a hyperplane to which we associate an outward normal $N_a(\sigma)$ in $S^3$, four out of which are linearly independent, and ten triangles from which we construct the ten simple bivectors $B_{ab}(\sigma)=({\bf b}_{ab}^-(\sigma),{\bf b}_{ab}^+(\sigma))$.

We introduce the diffeomorphism $\zeta: S^3 \rightarrow \SU(2)$; $N \mapsto \zeta(N) = N^0 1\!\!1 + i N^j  \sigma_j$ between the space of unit normals $S^3 \subset \R^4$, and the unitary group $\SU(2)$. From this map, we can relate  four-dimensional rotations $G$ to the action of the spin group $\Spin(4)$ on $\SU(2)$: $\zeta(G N) = X_- \zeta(N) X_{+}^{-1}$. Using this one can see that given the five normals $N_a(\sigma)$ one can pick five $\Spin(4)$ group elements $G_a=(X_a^{-}, X_a^{+})$ via the transitive action of $\Spin(4)$ on $S^3$:
$$
\zeta(N_a(\sigma)) = X_a^- \zeta(\pole) (X_a^+)^{-1}=X_a^- (X_a^+)^{-1}.
$$
These are not unique, different choices being related by the diagonal $\SU(2)$ subgroup which leaves $\pole$ invariant. This acts separately for each tetrahedron and is the action of $\SU(2)$ in the statement of the lemma.

Now pick one such set of values $\{X^\pm_a\}$. From the bivectors $B_{ab}(\sigma)$ and the group element $G_a$, construct the bivectors
$$
({\mathbf  c}_{ab}^-,{\mathbf c}_{ab}^+) := ((X_a^-)^{-1} , (X_a^+)^{-1})({\bf b}_{ab}^-(\sigma),{\bf b}_{ab}^+(\sigma)).
$$
From the discussion above, it is easy to see  that this bivector lies in a hyperplane orthogonal to $G_a^{-1} N_a
= \pole$. Accordingly, it necessarily has equal left and right components ${\mathbf c}_{ab}^- = {\mathbf c}_{ab}^+ := k_{ab} \, {\mathbf n}_{ab}$. (This is true for any such choice of
$\{X^\pm_a\}$.)  The closure and the orientation of the bivectors $B_{ab}(\sigma)$ respectively imply, by inverting the above equation, the critical point equations \eqref{Re} $\square$.

Note that the condition that the areas are half-integer is only required to meet the definition of boundary data. Thus the lemma would still be true if this condition were removed and the definition of boundary data did not require the areas to be half-integer.

We can now state the converse, which is the key to interpreting the critical point equations.

\begin{theo} {\bf (Reconstruction theorem)}
\label{simplex-bivector}
Given a fixed, non-degenerate boundary data $\mathcal{B}$ satisfying closure, $\sum_{b \neq a} k_{ab} \,\, {\mathbf n}_{ab} = 0$,
a set $X_{a}^{\pm}\in\SU(2)$, $a=1,...,5$, solving the orientation equations $X_a^{\pm} {\mathbf n}_{ab} = - X_b^{\pm } {\mathbf n}_{ba}$ is one of two geometric types:
\begin{itemize}
\item The solutions determine a non-degenerate geometric 4-simplex $\sigma$ in $\R^4$, defined up to translation and inversion, and a parameter $\mu=\pm1$, or
\item The solutions determine a single vector geometry.
\end{itemize}
In the first case, the bivectors $B_{ab}(\sigma)$ of the 4-simplex $\sigma$ satisfy
$$B_{ab}(\sigma)=\mu k_{ab}(X^-_a,X^+_a)(\mathbf n_{ab},\mathbf n_{ab}),$$
 for all $a\ne b$.
\end{theo}

{\em Proof.} The proof is based on the bivector geometry theorem. Consider the set of ten bivectors $B_{ab}$ constructed as follows:
$$
B_{ab} = (\mathbf b_{ab}^-,\mathbf b_{ab}^+) := k_{ab} \, (X_a^-,X_a^+)({\mathbf n}_{ab}, {\mathbf n}_{ab}).
$$
These satisfy the orientation, closure and diagonal simplicity bivector geometry conditions. The cross-simplicity condition follows from the fact that the bivector $({\mathbf n}_{ab}, {\mathbf n}_{ab})$ is orthogonal to the north pole $\pole=(1,0,0,0)$, and so the four bivectors $B_{ab}$ with fixed $a$ are orthogonal to the vector $N_a$ obtained by the action of $(X_a^-,X_a^+)$ on $\pole$.
The tetrahedron condition follows from the fact that
$\tr \, \left(B_{ab}[B_{ac},B_{ad}]\right)$ is proportional (by a non-zero constant) to plus or minus the squared volume of the geometric tetrahedron determined by the boundary data, which is non-degenerate by hypothesis.

We have therefore constructed bivectors satisfying all of the bivector geometry constraints except for the non-degeneracy condition. To investigate this final constraint we need the following lemma.

\begin{lem}\label{Lemma 2}
Given a boundary data set $\mathcal{B}$ satisfying \eqref{Re}, the $\SU(2)$ group elements $X_{a}^{\pm}$ satisfying \eqref{critical} are such that the corresponding constructed five unit vectors $N_a\in S^3$ fall into one and only one of the following two classes:
\begin{enumerate}
\item At least three out of the five $N_a$ are linearly independent
\item The $N_a$ are all proportional, i.e., $N_a =\pm N$ for some fixed $N$.
\end{enumerate}
In the first case the constructed bivectors satisfy the non-degeneracy condition of theorem \ref{bivectors}.
\end{lem}

To establish the first part of the lemma we show
\begin{itemize}
\item If three of the normals $\{N_a, N_b , N_c\}$ are proportional to $N$ , then consider a fourth normal $N_d$.
By the gluing condition the bivectors $B_{ea}, B_{eb}, B_{ec}$ lie in the hyperplane orthogonal to $N$ and so by closure so does $B_{ed}$. Thus the tetrahedron lies in this hyperplane and so $N_d$ is also proportional to $N$. The same holds for $N_e$.
\item In all the other possible cases, three of the normals $\{N_a, N_b , N_c\}$ must be pairwise linearly independent. Then one can conclude that all three together are a linearly independent set. This follows from the fact that the intersection of hyperplanes $a$ and $b$ and the intersection of hyperplanes $a$ and $c$ are non-parallel planes in hyperplane $a$ (due to the fact that they are faces of a non-degenerate tetrahedron).
\end{itemize}

Now, given three linearly independent normals to tetrahedra $\{N_a,N_b,N_c\}$ we will construct six linearly independent
bivectors from the tetrahedral bivectors. We take $x^1_{ab},x^2_{ab}$
 to generate the plane $N_a^\perp\cap N_b^\perp$ and from these two vectors we complete $N_a^\perp$ with $x_A$
 and $N_b^\perp$ with $x_B$, we do this in such a way that $x_B$ also completes $N_b^\perp\cap N_c^\perp$, which can always be done. Then
we construct the five linearly independent bivectors
 $$\{x^1_{ab}\wedge x^2_{ab}, ~x^1_{ab}\wedge x_A, ~x^1_{ab}\wedge x_B,~ x^2_{ab}\wedge x_A,~ x^2_{ab}\wedge x_B\}.$$ Each of these lie in
the hyperplane of a tetrahedron, and thus is a linear combination of the terahedral bivectors. Together with $x_A\wedge x_B$ they span the
whole space of bivectors.

Now we take the intersection $N_b^\perp\cap N_c^\perp$ generated by $x^1_{bc},x^2_{bc}$ and complete $N_c^\perp$
with $x_C$. We have that $x_C=\alpha_1 x^1_{ab}+ \alpha_2 x^2_{ab} + \alpha_3 x_B + \alpha_4 x_A$ since these span
the whole 4 dimensional space. Furthermore $\alpha_4\neq 0$ since otherwise $N_c^\perp$ and $N_b^\perp$ would be
spanned by the same vectors and $N_c \parallel N_b$. Furthermore $x^1_{bc}=\beta_1 x^1_{ab}+ \beta_2 x^2_{ab} +
\beta_3 x_B$ can be chosen such that $\beta_3$ is non zero, since $N_a^\perp\cap N_b^\perp\ne N_b^\perp\cap
N_c^\perp$, due to the non-degeneracy of the boundary data. Then the new bivector $x^1_{bc}\wedge x_C$ contains the nonzero term $\alpha_4 \beta_3 x_A \wedge x_B$
along with terms in the set of five bivectors above.
Therefore a complete set of six linearly independent bivectors is
generated by the tetrahedral ones, and Lemma 2 is proved $\square$

The above lemma implies that the set of constructed bivectors $B_{ab}$ are one of only two types. In one case they satisfy all bivector geometry constraints, including the non-degeneracy condition. Then using theorem \ref{bivectors}, we can conclude that there exists a parameter $\mu = \pm 1$ and a geometric $4$-simplex $\sigma$, up to translation and inversion, with bivectors $B_{ab}(\sigma)$ given by $B_{ab}(\sigma) = \mu B_{ab}$, determined by the corresponding solutions to the critical point equations.

If the constructed bivectors do not satisfy the non-degeneracy condition, they all lie in the same hyperplane and hence determine a vector geometry which is not a non-degenerate geometric $4$-simplex. This concludes the proof of the reconstruction theorem $\square$


\subsection{Classification and uniqueness of the solutions}

An important input in the asymptotic formula is the classification of the solutions to the critical point equations.

\subsubsection{Symmetries of the critical points}\label{symmetries}
To start with we need to consider their  symmetries. These include the symmetries of the action discussed in \ref{Action Symmetries}. Given a set of five $\SU(2)$ elements $\{U_a\}$ solving \eqref{Re} we immediately see that $\{\epsilon_a Y U_a\}$ with $Y \in \SU(2)$ and $\epsilon_a = \pm 1$ is also a solution. The sign ambiguity is due to the fact that two $\SU(2)$ elements differing by a sign act in the same fashion on three dimensional vectors. If two solutions are related by the symmetries we write $\{U^1_a\} \sim \{U^2_a\}$. If they are not related by symmetries we write $\{U^1_a\} \nsim \{U^2_a\}$ and call them distinct.

Given two solutions to the critical point equations we have two ways of forming $\Spin(4)$ elements $(X_a^+,X_a^-)$ out of them, that is $X_a^\pm = U_a^1,\; X_a^\mp = U_a^2$. Thus we also have
$$P\colon (X^-,X^+)\to (X^+,X^-)$$
as a symmetry. This is not a symmetry of the action but a symmetry of the solutions of the critical point equations.

The induced action of $\Spin(4)$ and $P$ on the normal vectors $N_a=\zeta^{-1} \bigl( X_a^- (X_a^+)^{-1} \bigr)$ are the action of $\SO(4)$ and the parity transformation
$$ P\in {\rm{O}}(4),~~P:(v_0,  {\bf v} )\mapsto(v_0, -{\bf v}) ~~\mbox{where}~~ (v_0,{\bf v})\in\left(\R \pole\right)\oplus \pole^\perp$$
respectively, acting on $N_a\in\R^4$.
Together, these generate the group ${\rm{O}}(4)$, the full set of linear isometries of four-dimensional Euclidean space.

We can now translate the result of  lemma \ref{Lemma 2} into a statement on the group elements of the solutions. Note that using the isomorphism between $S^3$ and $\SU(2)$ we can turn the results for the normals in the degenerate sector, $N_a = \pm N$, into $X_a^- (X_a^+)^{-1} = \epsilon_a Y$. Thus we have ${X_a^-} \sim {X_a^+}$. The converse clearly also holds. Since we showed in lemma \ref{Lemma 2} that the only other case is full non-degeneracy we also have that if the two sets are distinct these group elements define a non-degenerate 4-simplex up to inversion.

Bivectors register the $\SO(3) \times \SO(3)$ quotient of the $\Spin(4)$ symmetry. The parity transformation $P$ flips  the selfdual and antiselfdual sectors when acting on bivectors: $P( {\bf b}^-,{\bf b}^+)=({\bf b}^+,{\bf b}^-)$. This can be seen from the definition of the sectors \eqref{selfdual}:
$$P:\pm B^{0i}+ \frac{1}{2}   \epsilon^i_{\; jk} B^{jk}\mapsto \mp B^{0i}+ \frac{1}{2} \epsilon^i_{\; jk} B^{jk}.$$
The second term has two spatial indices and so does not change sign.

As bivectors do not register the full $\SU(2)$ symmetry there is a residual set of symmetries not acting on them given by the signs $\epsilon^\pm_a$. Part of this symmetry is however registered by the normals used in the reconstruction theorem. These transform as $X_a^- (X_a^+)^{-1} \rightarrow \epsilon^+_a \epsilon^-_a X_a^- (X_a^+)^{-1}$, and thus $N_a \rightarrow \epsilon^+_a \epsilon^-_a N_a$.

In the case of a non-degenerate 4-simplex, the reconstruction theorem gives a pair of 4-simplexes  $\sigma$, $\sigma'$ related by inversion. The symmetries can be used to turn the tetrahedron normals into ones which are all outward pointing for one of these and all inward pointing for the other. At such a critical point then $\sigma$ and $\sigma'$ are distinguished, and the one which has all normals outward pointing will be used as the canonical 4-simplex associated to the critical point. This 4-simplex is denoted $\sigma$ in the following.

The symmetries do not affect the 3-vectors $\mathbf n_{ab}$. However they do affect their relation to the 4-simplex $\sigma.$ For a given 3-simplex $a$ the inverse of the $\SO(4)$ transformation determined by $X_a^-,X_a^+$ maps this face of $\sigma$ to  a tetrahedron $t'^a$ in $\R^3$ (the hyperplane orthogonal to $\pole$). The $\mathbf n_{ab}$ are either all outward-pointing normals for the faces of $t'^a$ , or they are all inward-pointing normals.
The orientation condition implies that if they are outward (inward)-pointing in one tetrahedron, they are outward (inward)-pointing, respectively, in all the others.
In fact, a calculation shows that the parameter $\mu=\pm1$ of theorem \ref{bivectors} for the 4-simplex takes the value $+1$ for outward-pointing tetrahedron normals and $-1$ for inward-pointing.

The action of $P$ flips $\mu\to -\mu$, whereas the action of $\Spin(4)$ leaves it unchanged. This follows from the fact that $P B_{ab}(\sigma) = - B_{ab}(P\sigma)$, as can be seen from the presence of the Hodge star in equation \eqref{geometric bivectors}.
Thus $\mu$ measures the orientation of the 4-simplex; $\mu=1$ if the orientation agrees with the orientation of its boundary specified in section \ref{Regge-like}, and $\mu=-1$ if it has the opposite orientation.

\subsubsection{Classification}\label{class}
We can now prove the following lemma:

\begin{lem}
Let $\mathcal{B}$ denote the set of fixed boundary data $\{{\mathbf  n}_{ab},k_{ab}\}_{a\ne b=1,...,5}$, with ${\mathbf n}$ in $S^2$, and $k \in \N /2$ satisfying closure
\be
\label{boundary}
\sum_{b\colon b \neq a} k_{ab} \,\, {\mathbf n}_{ab} = 0,
\ee
for all $a=1,...,5$. Consider a set of five $\SU(2)$ group elements $U_a$ solving the equations
\be
\label{equations}
U_a {\mathbf n}_{ab} = - U_b {\mathbf n}_{ba}.
\ee
There are at most two distinct sets of $U_a$.
\end{lem}

{\itshape Proof.} Assume there are three distinct  solutions to \eqref{equations}, say $\{U_a^1\}, \{U_a^2\}, \{U_a^0\}$. Then we can form non-degenerate solutions to the critical equations, e.g. $(X_a^{-},X_a^+) = (U_a^1, U_a^0)$ and $(X_a^{-},X_a^+) = (U_a^2, U_a^0)$. These then give two non degenerate geometric 4-simplices with the same boundary data up to inversion. The boundary data however gives the full set of tetrahedra forming the boundary of the geometric 4-simplex. However any two $n$-simplices with congruent $(n-1)$-faces are in fact congruent themselves (as discussed in section \ref{asymptotic}). Therefore the two 4-simplices must be the same up to rigid motion.   In particular they are related by $P$ and $\SO(3) \times \SO(3)$ up to inversion. Thus we have either that $\{U_a^1\} \sim \{U_a^2\}$, or, if the rigid motion mapping them to each other involves a parity, that $\{U_a^1\} \sim \{U_a^0\}$ and $\{U_a^0\} \sim \{U_a^2\}$. Both are in contradiction to our assumption that $\{U_a^1\}, \{U_a^2\}, \{U_a^0\}$ are distinct. It follows that we can never have three or more distinct solutions to \eqref{equations}.$\square$

From this it follows that there are three types of $\mathcal{B}$, distinguished by how many solutions $\{U_a\}$ to \eqref{equations} they allow:
\begin{enumerate}
\item The set $\mathcal{B}$ is such that there are no solutions to \eqref{equations}. This means the tetrahedra do not glue at all.
\item There is only one distinct solution to \eqref{equations}, thus there is only one set of solutions to the critical point equations. This is given by $(X_a^{-},X_a^+) = (U_a, U_a)$.
\item \label{geo} There are two distinct solutions to \eqref{equations} and we have, up to the symmetries of the action, four solutions to the critical point equations: $$(X_a^{-},X_a^+) \in \{  (U_a^{-}, U_a^+),(U_a^{+}, U_a^-), (U_a^{+}, U_a^+), (U_a^{-}, U_a^-)\}$$ These are in one to one correspondence to non degenerate geometric 4-simplices up to rigid motion.
\end{enumerate}

\subsection{The action at critical points}\label{ReggeSection}

In this section the action at a critical point is expressed in terms of the underlying geometry.

The critical points satisfy closure and the conditions \be \label{critical1} X^{\pm}_{ab} ( {\mathbf n}_{ba}) =
-{\mathbf n}_{ab}, \ee for all $a\ne b$. The lift of this equation to the coherent states involves a phase \be
\label{critcoherent} X^{\pm}_{ab} | \mathbf n_{ba} \ra = e^{i \phi_{ab}^{\pm}} | -\mathbf n_{ab} \ra. \ee
The phases $\phi_{ab}^+$ and $\phi_{ab}^-$ are defined modulo $2\pi$. However it is the sum and difference of the phases that enters into the geometrical expressions.

Evaluating the action $S=iA$, either \eqref{action} or \eqref{gamma > 1 action}, on the critical points leads to
\be \label{criticalaction} A = \sum_{a < b} k_{ab} \, (
\phi_{ab}^{+} + \phi_{ab}^{-} ) + \gamma \, k_{ab} \, ( \phi_{ab}^{+} - \phi_{ab}^{-} ), \ee for all $\gamma\ne1$,
where we have used the corresponding relation between $j^{\pm}$ and $k$. In this formula, the phases enter via the variables $\phi_{ab}^{+} + \phi_{ab}^{-}$ and $\phi_{ab}^{+} - \phi_{ab}^{-}$ which are both defined mod $4\pi$, but with one remaining indeterminacy given by the simultaneous replacement
$$\phi_{ab}^{+} + \phi_{ab}^{-}\to \phi_{ab}^{+} + \phi_{ab}^{-}+2\pi$$
$$\phi_{ab}^{+} - \phi_{ab}^{-}\to \phi_{ab}^{+} - \phi_{ab}^{-}+2\pi.$$
Taking into account the fact that the $k_{ab}$ are half-integers and that $(1+\gamma)k_{ab}$ are integers, one can see explicitly that $A$ is well-defined mod $2\pi$. The indeterminacies may be fixed by choosing suitable coordinate ranges; we will use
$$-2\pi\le \phi_{ab}^{+} + \phi_{ab}^{-}<2\pi,$$
$$-\pi\le\phi_{ab}^{+} - \phi_{ab}^{-}<\pi.$$

Further analysis of the action is done by separating two special cases.

\subsubsection{Non-degenerate 4-simplex, $X^+$ and $X^-$ distinct} \label{Sec-Dihedral angles}

In this case, the critical point is given by the sets $\{X^+_a\}$ and $\{X^-_a\}$ which are assumed to be distinct from each other, i.e., not related by a symmetry. According to theorem \ref{simplex-bivector}, these correspond to a pair of geometrical 4-simplexes $\sigma$ and $\sigma'$, related by inversion. The results of this section will show that the phase difference term in the action \eqref{criticalaction}  evaluates to the Regge action of the geometric 4-simplex. Further, in the case of a Regge state it will be shown that the phase sum term in the action vanishes.

According to the discussion in \ref{symmetries}, there is a discrete symmetry of the action which relates any critical point to one in which $N_a\in\R^4$ is the {\em outward} pointing normal of the 4-simplex $\sigma$ for each tetrahedron $a$. Therefore it is enough to evaluate the action on these critical points.


The Regge action for a flat Euclidean four-simplex is a boundary term that involves the dihedral
angles $0<\Theta<\pi$ between the normals of the boundary tetrahedra \cite{barrett-1999-3}. This angle is
calculated through the scalar product between the  outward normals to the tetrahedra \bea \label{dihedral}
\cos \Theta_{ab} &=& N_a\cdot N_b\nn \\
                &=& \frac{1}{2} \, \tr (\zeta(N_a) \zeta(N_b)^{-1}) \nn \\
                &=& \frac{1}{2} \, \tr (X_{ab}^- (X_{ab}^{+})^{-1}),
\eea
where we have used the fact that we are considering only the critical points corresponding to outward pointing constructed normals, that is, $N_a(\sigma) = N_a$ for all $a$.
The key point is that the above angle can be related to the rotation angle associated to the rotation
stabilising the vector ${\mathbf n}_{ab}$, since the coupling of  two of the equations \eqref{critical1} leads to
the relation
\be \label{isotropy} X_{ab}^- (X_{ab}^{+})^{-1} {\mathbf n}_{ab} = {\mathbf n}_{ab}.  \ee
Similarly,
\eqref{critcoherent} leads to
\be\label{isotropy2} X_{ab}^- (X_{ab}^{+})^{-1}| \mathbf n_{ab}\ra= e^{i(\phi_{ab}^+
- \phi_{ab}^-)}|\mathbf n_{ab}\ra \ee
This shows that the $\SU(2)$ matrix $X_{ab}^- (X_{ab}^{+})^{-1}$ generates
rotations about the $\mathbf n_{ab}$ axis, and has eigenvalues $e^{i(\phi_{ab}^+ - \phi_{ab}^-)}$ and its complex
conjugate. It can be written
\be \label{phase equation} X_{ab}^- (X_{ab}^{+})^{-1}=\exp i(\phi_{ab}^+ - \phi_{ab}^-){\pmb\sigma}\cdot \mathbf n_{ab}=\exp 2(\phi_{ab}^+ - \phi_{ab}^-)\mathbf L\cdot \mathbf n_{ab}
\ee
using $\mathbf L=\frac i2{\pmb\sigma}$, as before.

 Equation \eqref{dihedral}
then shows that
$$ | \phi_{ab}^{+} - \phi_{ab}^{-}| = \Theta_{ab}.$$
This argument does not resolve the sign of the left-hand side, and a more delicate argument is required. The core
of our argument is that this sign depends only on the relative orientation of the bivectors $*B_{ab}$ and the
geometric 4-simplex, in fact:

\begin{lem}
\label{dih}
The relative sign between the angle $\phi_{ab}^{+} - \phi_{ab}^{-}$ and the dihedral angle $\Theta_{ab}$ is determined by the parameter $\mu=\pm1$ controlling the relative sign between the geometrical and constructed bivectors of theorem \ref{simplex-bivector}:
$$ \phi_{ab}^+ - \phi_{ab}^- = \mu \Theta_{ab}.$$
\end{lem}

{\em Proof.} The proof is based on the explicit construction of the dihedral rotation associated to the dihedral angle.
Recall that a simple bivector $B_{ab}$ admits two stabilising $\Spin(4)$ subgroups. This is due to the isomorphism $\Lambda^2(\R^4) \cong \spin(4)$ and to the fact that $\spin(4)$ admits a two-dimensional Cartan subalgebra. This is reflected in the fact that $B_{ab}$, regarded as a Lie algebra element, obviously commutes with itself but also with its Hodge dual $*B_{ab}$. Hence, the $\Spin(4)$ subgroups $G_{ab}$ and $G_{ab}^{\bot}$ generated respectively by the Lie algebra elements $B_{ab} / | B_{ab} |$ and $*B_{ab} / | B_{ab} |$ stabilise $B_{ab}$. Geometrically, the subgroups $G_{ab}$ and $G_{ab}^{\bot}$ correspond to simple rotations acting in the plane defined by $B_{ab}$ and in its orthogonal complement respectively.

Consider a geometrical $4$-simplex $\sigma$ and a particular bivector $B_{ab}(\sigma)$. The corresponding isotropy group $G_{ab}^{\bot}$ contains an element of particular interest; the dihedral rotation $\widehat D_{ab}\in\SO(4)$ mapping the
normal $N_a$ to the normal $N_b$ and stabilising the orthogonal plane
\be
\label{dihedral-simplex}
\widehat D_{ab} =  \exp \left(\Theta_{ab} \frac{N_b\wedge N_a}{|N_b\wedge N_a|} \right).
\ee
This is clearly a simple rotation in the plane spanned by $N_a$ and $N_b$, so we just need to check that  the rotation has the right direction and scale factor. To
do this we expand the exponential above to first order and apply it to $N_a$
obtaining
\be
\left(1\!\! 1 +\Theta_{ab}\frac{N_b\wedge N_a}{|N_b\wedge N_a|}\right)\cdot
N_a=N_a+\frac{\Theta_{ab}(N_b-N_a\cos{\Theta_{ab}})}{|\sin{\Theta_{ab}}|}
\ee
where we have used \eqref{bivnorm} and \eqref{bivaction}. For small positive $\Theta_{ab}$, clearly the expression above reads
$$
\widehat D_{ab} \cdot N_a \cong N_b,
$$
which means \eqref{dihedral-simplex} is correct.

Using the definition \eqref{geometric bivectors}, theorem \ref{simplex-bivector} and the isomorphisms of section
\ref{bivectorsection},  the bivector is expressed in terms of the boundary data as
$$\frac{N_a\wedge N_b}{|N_b\wedge N_a|}=-\mu (X^-_a,X^+_a)(\mathbf n_{ab},-\mathbf n_{ab})
=(-\mu X^-_a\mathbf n_{ab},\mu X^+_a\mathbf n_{ab}).$$
Introduce the exponential
$$d_{ab}=\pm\exp{\mu\Theta_{ab}\mathbf n_{ab}\cdot\mathbf L}\in\SU(2).$$
A lift of the dihedral rotation to $\SU(2)\times\SU(2)$ is
\begin{equation}
\label{dihedral-simplex2}
D_{ab}
= (X^-_a\rhd d_{ab},
X^+_a\rhd d_{ab}^{-1})
\end{equation}
where the symbol `$\rhd$'
denotes the adjoint action of $\SU(2)$ on itself, $X\rhd d=XdX^{-1}$. Either choice of sign in $d_{ab}$ gives a lift, reflecting the 2-1 ambiguity in the spin group. The choice of sign is arbitrary for the purposes of this proof.

Finally, the equation $\widehat D_{ab}N_a=N_b$ implies that
\begin{multline}\zeta(\widehat D_{ab}N_a)=\left(X^-_a\rhd\exp{\mu\Theta_{ab}\mathbf n_{ab}\cdot\mathbf L}\right)
X_a^-(X_a^+)^{-1}
\left(
X^+_a\rhd\exp{-\mu\Theta_{ab}\mathbf n_{ab}\cdot\mathbf L}
\right)^{-1}\\
=X_a^-\left(
\exp{2\mu\Theta_{ab}\mathbf n_{ab}\cdot\mathbf L}
\right)
(X_a^+)^{-1}=X_b^-(X_b^+)^{-1}
\end{multline}
Using \eqref{phase equation} now gives
$$\exp{2\mu\Theta_{ab}\mathbf n_{ab}\cdot\mathbf L}=\exp 2(\phi_{ab}^+ - \phi_{ab}^-) \mathbf n_{ab}\cdot\mathbf L$$
and hence $\mu\Theta_{ab}=\phi_{ab}^+ - \phi_{ab}^-$.  This concludes the proof of lemma \ref{dih}  $\square$ .


In rest of this section it is shown how the action \eqref{criticalaction} simplifies for the case of a Regge state.
In \eqref{dihedral-simplex2} a key role is played by the $\SU(2)$ element
\be\label{kappa-definition} d_{ab}=\kappa_{ab}\exp{\mu\Theta_{ab}\mathbf n_{ab}\cdot\mathbf L},\ee
with $\kappa_{ab}=\pm1$. By \eqref{phase equation} we have
\be
\label{dsquare}
d_{ab}^2=X_{ab}^-\left(X_{ab}^+\right)^{-1}
\ee
  Let us now define
  \be
  h_{ab}^+ :=
d_{ab} X_{ab}^+ ~~~~~\mbox{and~~~~~} h_{ab}^-:= d_{ab}^{-1} X_{ab}^-
\ee
We can check in fact that
$h_{ab}^+=h_{ab}^-=:h_{ab}$ by calculating $(h_{ab}^-)^{-1}h_{ab}^+=(h_{ab}^+)^{-1}h_{ab}^-=1\!\! 1$.

The key point now is that the rotation $(X^-_{ab}, X^+_{ab})$, that maps $\pole^{\bot}$ into $\pole^{\bot}$, can
then be uniquely decomposed into two simple rotations:
\be \label{rotation} (X^-_{ab}, X^+_{ab}) =
 (d_{ab} , d_{ab}^{-1})\,(h_{ab},h_{ab}),
\ee
The element $(h_{ab},h_{ab})$ belongs to the diagonal $\SU(2)$ subgroup of
$\Spin(4)$ leaving $\pole$ invariant and is thus a rotation in the hyperplane $\pole^{\bot} \cong \R^3$. This
three-dimensional rotation glues the triangle corresponding to the bivector $({\mathbf n}_{ba}, {\mathbf n}_{ba})$
onto the triangle associated to the bivector $({\mathbf n}_{ab}, {\mathbf n}_{ab})$. Note that by
\eqref{critcoherent}, we have:
\bea\label{gluing}
d^{-1}_{ab} h_{ab} |\mathbf n_{ba}\rangle &=& X_{ab}^+ |\mathbf n_{ba}\rangle = e^{i \phi_{ab}^{+}} J| \mathbf n_{ab} \ra\nn\\
d_{ab}       h_{ab}|\mathbf n_{ba}\rangle &=& X_{ab}^-|\mathbf n_{ba}\rangle  = e^{i \phi_{ab}^{-}} J| \mathbf n_{ab}
\ra\eea

Now consider the following diagram:
\begin{equation}\label{commutative}
\xymatrix{\ar @{} [dr]  t_a \ar[d]_{(g_{ab},g_{ab})} \ar[rr]^{(X_a^-,X_a^+)} && ~\tau_a \ar[d]^{D_{ab}}   \\
t_b\ar[rr]_{(X_b^-,X_b^+)} && ~\tau_b  }
\end{equation}
where $t_a\in \R^3\subset\R^4$ are the tetrahedra at the north pole (i.e. in the hyperplane orthogonal to $\pole$),
$\tau_a\in\R^4$ are the actual geometrical ones in the 4-simplex $\sigma$ and the $g_{ab}\in \SU(2)$, defined in
\eqref{g_ab}, are the maps that effectively glue the triangles $\phi_a(\Delta_{ab})$ and $\phi_b(\Delta_{ab})$,
and leave the normals inside $t_a$ and $t_b$ anti-parallel. Note that by the reconstruction theorem, the maps in the diagram
commutes when acting on both the triangles $\phi_a(\Delta_{ab})$ and on the internal normals $\mathbf n_{ab}$. Since furthermore all
maps in the diagram are orientation preserving, the composition maps coincide on a three dimensional subspace and
are orientation preserving, which means the $\SO(4)$ action of the maps in the diagram commutes. This means the diagram of maps as elements of $\Spin(4)$ is commutative up to sign. Therefore it is possible to now determine the sign $\kappa_{ab}$ in the definition of $d_{ab}$ as the one that makes this diagram of maps commutative in $\Spin(4)$.

Hence we have that
$$(X_a^- d_{ab}, X_a^+ d_{ab}^{-1})= (X_b^-g_{ab},X_b^+g_{ab})$$
which gives
$$g_{ab} = (X_b^-)^{-1} X_a^-  d_{ab}=(X_b^+)^{-1} X_a^+ d_{ab}^{-1},$$
and which also tells us that $h_{ab} = g_{ab}^{-1}$.  If we substitute this fact in \eqref{gluing} and use the action of $g_{ab}^{-1}$ on the coherent state $| \mathbf n_{ba} \ra$, then we obtain
 \bea
 d_{ab}^{-1}  J| \mathbf n_{ab} \ra &=& e^{i\phi_{ab}^{+}} J| \mathbf n_{ab} \ra\nn\\
d_{ab}J| \mathbf n_{ab} \ra         &=& e^{i \phi_{ab}^{-}} J| \mathbf n_{ab} \ra
\eea
Taking the eigenvalues, and writing $\kappa_{ab}=(-1)^{\nu_{ab}}$ with $\nu_{ab}=0\text{ or }1$ gives
$$\phi_{ab}^+=\frac 12\mu\Theta_{ab}+\pi\nu_{ab}$$
$$\phi_{ab}^-=-\frac12\mu\Theta_{ab}+\pi\nu_{ab}$$
mod $2\pi$. Adding these gives
$$\phi_{ab}^++\phi_{ab}^-=2\pi\nu_{ab}$$
as an angle between $0$ and $4\pi$.

The signs $\kappa_{ab}$ cannot be always equal to one, since the symmetries $g_{ab}\to \epsilon_a\epsilon_b g_{ab}$, for $\epsilon_a=\pm1$, change them by
$$\kappa_{ab}\mapsto \epsilon_a\epsilon_b\kappa_{ab}.$$
Exactly the same transformation also occurs as a result of applying the symmetries
$(X^-_{a},X_{a}^+) \mapsto(\epsilon_a X^-_{a},\epsilon_a X_{a}^+)$.

However it is possible to show that these symmetries can be used to transform either the $g$ or the $X$ so that $\kappa_{ab}=1$ for all $a$ and $b$.

\begin{lem}\label{pifactors}
The signs defined in \eqref{kappa-definition} obey
$\kappa_{ab}=\epsilon_a\epsilon_b$ for some $\epsilon_a=\pm1$.
\end{lem}

{\em Proof.}
The geometrical interpretation of the dihedral rotation $\widehat D_{ab}\in\SO(4)$ is that it restricts to the Levi-Civita connection $\omega_{ab}$ on the boundary $\partial \sigma$. Let $\tau_a$ be the $a$-th tetrahedral face of $\sigma$. Then the dihedral rotation maps the tangent spaces
$$\widehat D_{ab}\colon T \tau_a\to T\tau_b.$$
Therefore the lift $D_{ab}\in\Spin(4)$ represents the spin connection in some gauge; indeed according to \eqref{commutative} it is gauge-equivalent to the $g_{ab}$, which does indeed represent the spin connection on the boundary.

According to the discussion in section \ref{Regge states}, the holonomy of the spin connection around an edge in the boundary 3-manifold $\partial\sigma$ is uniquely determined; it is a rotation by the deficit angle of the Regge geometry, modulo $4\pi$. This can be characterised as the rotation which deforms continuously to the identity rotation (and not a $2\pi$ rotation) when the geometry around the edge is deformed to the flat geometry with zero deficit angle. Alternatively, one can view this as a statement about the contraction of the loop to a point in a smoothing of the geometry.

The Levi-Civita connection has a canonical lift, given by
\be  \widetilde d_{ab}=\exp{\mu\Theta_{ab}\mathbf n_{ab}\cdot\mathbf L}=\kappa_{ab}d_{ab},\ee
which is used to give the canonical lift of the dihedral rotation,
\begin{equation}
\widetilde D_{ab}
= (X^-_a\rhd \widetilde d_{ab},
X^+_a\rhd \widetilde d_{ab}^{-1}).
\end{equation}
The holonomy around an edge of the boundary which lies in the three tetrahedra $a$, $b$ and $c$ is
$$ \widetilde D_{ca}\widetilde D_{bc}\widetilde D_{ab}.$$
This is the correct holonomy because it lifts the Levi-Civita connection and reduces to the identity element of $\Spin(4)$ if the dihedral angles $\Theta_{ab}$, $\Theta_{bc}$, $\Theta_{ca}$ are deformed continuously to zero. Therefore this holonomy is equal to
$$  D_{ca} D_{bc}D_{ab}.$$
 Thus around each edge one has the cocyle condition
$$\kappa_{ca}\kappa_{bc}\kappa_{ab}=1,$$
which, since $H^1(S^3,Z_2)=0$, implies that $\kappa_{ab}=\epsilon_a\epsilon_b$ for some values of $\epsilon_a=\pm1$. $\square$
\medskip

Finally, putting these results in the action \eqref{criticalaction} results in
\be\label{nondegenerateaction} A=\mu\gamma\sum_{a<b} k_{ab}\Theta_{ab}.\ee

\subsubsection{Non-degenerate 4-simplex, $X^+$ and $X^-$ related by symmetry} \label{weird terms}

In this case $\{X^+_a\}\sim\{X^-_a\}$, which implies that
$\phi^+_{ab}=\phi^-_{ab}$. Hence the action is
$$A=\sum_{a<b} 2 k_{ab} \phi^+_{ab}.$$
According to section \ref{class}, for a non-degenerate 4-simplex there exist two distinct solutions to the critical equations, $\{U^+_a\}$ and $\{U^-_a\}$. The case here arises when $\{X^+_a\}\sim\{X^-_a\}\sim \{U^+_a\}$, for example.
Putting $X'^+_a=X^+_a$ and $X'^-_a=U^-_a$, then the reconstruction theorem applies to the data $\{X'^+_a\}, \{X'^-_a\}$, as in the previous section.
Then the results of the previous section applied to
$\{X'^+_a\}, \{X'^-_a\}$ show that
$$\phi^+_{ab}=\frac 12\mu\Theta_{ab}+\pi\nu_{ab}.$$
The parameters $\mu$ and $\Theta_{ab}$ are those determined by $\{X'^+_a\}, \{X'^-_a\}$.

Thus the action is
\be\label{degenerateaction}
A=\mu\sum_{a<b}  k_{ab}\Theta_{ab},
\ee
the sum over the $\nu_{ab}$ terms again vanishing by lemma \ref{pifactors}.

\subsection{Proof of theorem \ref{big theorem new}}

The results of the previous sections can now be assembled into the proof of the main theorem.
The idea of the proof is to apply the method of extended stationary phase to the integral formula \eqref{amplitude2} for $\gamma<1$ and \eqref{amplitude3} for $\gamma>1$. However
before this is done, it is necessary to account for the symmetries of the formula.

According to the results of section \ref{class}, given fixed boundary data, the number of distinct classes of critical points not related by the symmetries of the action (section \ref{Symmetries of the action}) is either none, one or four solutions.

For any critical point, the discrete symmetries give $2^{10}$ critical points which give the same contribution to the asymptotic formula. However four of these discrete symmetries (with $\epsilon^\pm_a$ independent of $a$) are in the class of continuous symmetries. Thus in each symmetry class there are $2^{8}$ connected manifolds of critical points, each of which has the free action of the $\Spin(4)$ continuous symmetries. In fact in the case of a non-degenerate 4-simplex, a group of order $2^4$ was used to make the normals $N_a$ outward-pointing, and another group of order $2^4$ was used to set $\kappa_{ab}=1$. However it is actually not necessary to keep track of this: the contribution of each of these connected components is the same in each symmetry class and can be evaluated by looking at any one of them.

 In order to apply the method of extended stationary phase  we need to ensure that the stationary points are isolated.
 This is accomplished by the change of variables $\tilde{X^\pm}_a =   (X_5^\pm)^{-1} X_a^\pm$. After this $X^\pm_5$ drops out of the action, and we are left with an $\SU(2)\times\SU(2)$ integration which is redundant.
The `gauge fixed' integral formulas then have isolated critical points related only by the discrete symmetries and can now be evaluated using extended stationary phase.

Consider first the case of four solutions. This is exactly if the boundary data corresponds to the boundary of a non-degenerate geometric 4 simplex.  For this boundary data, consider the symmetry class of critical points with distinct ${X^+}$ and ${X^-}$.  In section \ref{Sec-Dihedral angles} it was shown that the action at these points evaluates to \eqref{nondegenerateaction}. From section \ref{class} we know that exactly two such solutions occur and that they are related by parity. From section \ref{symmetries} we know that parity only changes the sign of $\mu$ in \eqref{nondegenerateaction}. We can express the asymptotic result in terms of the solution $\{X^-,X^+\}=\{U^-,U^+\}$ which has $\mu = -1$. The phase of this asymptotic contribution to the amplitude is then given by $$A_{-+}=-\gamma\sum_{a<b} k_{ab}\Theta_{ab}$$ with $\Theta_{ab}$ the dihedral angles of geometric 4-simplex corresponding to the boundary. The contribution from $\{U^+,U^-\}$ has phase $$A_{+-}=\gamma\sum_{a<b} k_{ab}\Theta_{ab}.$$

The two additional terms we obtain in section \ref{class} correspond to the solutions $\{U^+, U^+\}$ and $\{U^-, U^-\}$. In section \ref{weird terms} we evaluated the action at these points as \eqref{degenerateaction}. Consider the $\{U^+,U^+\}$ solution. The $\mu$ occurring in the evaluation of its action is the same as in the evaluation of $\{U^+,U^-\}$. Therefore we have $\mu = 1$ for this point and $\mu = -1$ for $\{U^-,U^-\}$. The phases of their contribution to the asymptotics are therefore $$A_{++}=\sum_{a<b} k_{ab}\Theta_{ab}$$ and $$A_{--}=-\sum_{a<b} k_{ab}\Theta_{ab}$$ respectively.

As the final element of the asymptotic contributions of each of the solutions we need to calculate their Hessian matrices which are given by

\be H_{cd}^{ij, \pm \pm} = \left(\frac{\partial^2S}{\partial X^{i,
\pm}_c\partial X^{j, \pm}_d}\right) \ee
for $\gamma < 1$. This is now a 24 by 24 matrix as we have factored out the integration with respect to $X_5^\pm$ and $c$ and $d$ range from 1 to 4 only. We write $H(\{X^+,X^-\})$ for the Hessian evaluated at the critical point $\{X^+,X^-\}$. The Hessian at the critical points is given explicitly in terms of geometric data corresponding to these points in section \ref{Hessian section}. The expression for $\gamma > 1$ is only marginally more complicated and is given there as well. In both cases we have $\det(H(\{X^+,X^-\})) = \det(H(\{X^-,X^+\}))$ (see next section). This allows us to combine the asymptotic contributions from the two solutions with distinct elements into one cosine. 
The factor $a(x_0)$ in the stationary phase formula is the normalising factor for the measure on $8$ copies of $\SU(2)$ in coordinates which agree with the derivatives; each copy is a factor of $(4\pi)^{-2}$.
  
As each of the critical solutions occurs $2^{8}$ times  we thus find that the complete asymptotic behavior of \eqref{amplitude2} is given by:
\begin{multline}
f_4(\psi_\lambda )
\sim
(-1)^{\chi'}
\left(\frac{2 \pi}{\lambda}\right)^{12}      \frac{2^8 }{ (4\pi)^{16}  }  \left(
2 \frac{\cos ( \lambda \gamma \sum_{a<b} k_{ab} \Theta_{ab})}{\sqrt{\det H(\{U^+, U^-\})}} \right.\,
\nn \\  
+ \frac{\exp ( i \lambda \sum_{a<b} k_{ab} \Theta_{ab})}{\sqrt{\det H(\{U^+, U^+\})}} \,
+
\left.\frac{\exp ( - i \lambda \sum_{a<b} k_{ab} \Theta_{ab})}{\sqrt{\det H(\{U^-, U^-\})}}   \right)\, .
\end{multline}
This establishes the non degenerate boundary data with $\gamma < 1$ part of theorem \ref{big theorem new} with
$$
N^\gamma_{\epsilon_1 \epsilon_2} =   \frac{2^8 }{ (4\pi)^{16}  } (\det H(\{X^{\epsilon_1}, X^{\epsilon_2}\}))^{-\frac{1}{2}}
$$
and $\epsilon_1 , \epsilon_2 = \pm$.

 For $\gamma > 1$ we have four additional variables, ${\bf m}_{ab}$, ${\bf m}_{ba}$ per propagator in the action. Furthermore the resolution of the identity contributes a factor of $(d_{j_{ab}^+})^2 = (2j_{ab}^+ + 1)^2$. Therefore we get an overall additional factor of
$$
 (-1)^{(\gamma - 1)k_{ab}}   \left(\frac{2 \pi (\lambda 2 (\gamma + 1) k_{ab} + 1)}{ 4 \pi   \lambda}\right)^2 \sim (-1)^{(\gamma - 1)k_{ab}}  ( (\gamma + 1) k_{ab})^2
$$
per propagator. Thus the numerical factors for $\gamma > 1$ are 
$$
N^\gamma_{\epsilon_1 \epsilon_2} = 
   \frac{2^8 }{ (4\pi)^{16}  } (\det H(\{X^{\epsilon_1}, X^{\epsilon_2}\}))^{-\frac{1}{2}} \prod_{a<b}
 (-1)^{(\gamma - 1)k_{ab}} ( (\gamma + 1) k_{ab})^2 .
$$

For a set of boundary data which is not a non-degenerate 4-geometry but forms a vector geometry, we only have one contribution to the asymptotic formula up to symmetries. For this case we did not evaluate the phase of the single contribution. The stationary phase formula gives
\bea
|f_4( \psi_\lambda )|
&\sim& \left(\frac{2 \pi}{\lambda}\right)^{12} \frac{2^8 }{ (4\pi)^{16}  }
\frac{1}{\sqrt{| \det H(\{X, X\})|}}
\eea
for $\gamma < 1$ and the same prefactors as in the first case for $\gamma > 1$.

For a set of boundary data that is neither a non-degenerate 4-geometry nor a vector geometry,
there are no stationary points of the action and the stationary phase formula states that $f_4$ decreases faster than any power of $\lambda$.
\be
f_4( \psi_\lambda  ) = o(\lambda^{-N}) \; \; \; \forall N
\ee

This concludes the proof of theorem \ref{big theorem new} apart from the facts used about the Hessian which are established in the next section.$\square$

\subsection{The Hessian}
\label{Hessian section}
This section provides an explicit formula for the Hessian matrix which appears in the asymptotic formula. The symmetry of the Hessian under the interchange of $+$ and $-$ is then apparent.
 We will begin by discussing the ${\gamma < 1}$ case. Here the Hessian matrix of the full action is
a $24 \times 24$ matrix defined as
\be H_{cd}^{ij, \pm \pm} = \left(\frac{\partial^2S}{\partial X^{i,
\pm}_c\partial X^{j, \pm}_d}\right)
\ee
As the action is a sum of left and right sector, the terms off diagonal
with respect to the $\pm$ index are $0$ and $H$ is a block diagonal matrix
\be
H = \left(
\begin{array}{cc}
    H^{--}               &    0   \\
            0       &   H^{++}    \\
\end{array}
\right) \ee where $H^{++},H^{--}  $ are the $12 \times 12$  Hessian matrices for the left and right sectors
respectively.


As the Hessian is block diagonal, its determinant can be written as the product of the determinants of the two
submatrices
\be
\det ( H  ) = \det (H^{--})   \det (H^{++} ).
\ee
For clarity we just give the result for $\det (H^{++})$, $\det (H^{--} )$ is the same. From \eqref{action}, using again
\eqref{variation of X} for the variation of $\SU (2)$ elements we obtain for the elements diagonal in the vertex
index $c$:
\be
H^{ij,++}_{cc} =
\frac{\partial^2S}{\partial X^{i,+}_c\partial X^{j,+}_c}
=-\frac{1}{4} \sum_{b\neq
c}2j^+_{cb}\left(\delta^{ij}-i\epsilon^{ijk}V^{k,+}_{cb}-V^{i+}_{cb}V^{j,+}_{cb}\right)
\ee
Where $V^{k,+}_{cb}$ was defined in \eqref{vector}.
Now if we denote a set of critical points by $\{ X^-, X^+ \}$,
and use Lemma \ref{spinor-vector} to evaluate the Hessian at the critical points, we have
\be
H^{ij,++}_{cc}(X^+)=
\left. \left(\frac{\partial^2S}{\partial X^{i,+}_c\partial X^{j,+}_c}\right) \right|_{\{ X^+\}}=
-\frac{1}{4}\sum_{b\neq c}2j^+_{cb}\left(\delta^{ij}-n^{i,+}_{cb}n^{j,+}_{cb}\right)
\ee
which  is equivalent to the Hessian for the asymptotics of a coherent tetrahedron that was derived in \cite{livine-2007-76}. For the off-diagonal terms we obtain:
\be
H^{ij,++}_{cd}(X^+) =
\left.\left(\frac{\partial^2S}{\partial X^{i,+}_c\partial
X^{j,+}_d}\right)\right|_{\{ X^+\}}
=-\frac{1}{4} 2j^+_{cd}\left(\delta^{ij}-i\epsilon^{ijk} n^{k,+}_{cd}
-n^{i,+}_{cd}n^{j,+}_{cd}\right)
\ee
For $\gamma > 1$ there is an additional dependence on the coherent state
vectors $\mathbf m_{ab}$. This means the Hessian, while still symmetric in the left and right sectors, is now no
longer block diagonal. 


\section{Remarks and Conclusions}\label{Conclusion}
We have studied the semi-classical limit of the four-simplex amplitude of the Euclidean EPRL model for both sectors of the Immirzi parameter. Using the method of coherent states, we have shown that non-trivial contributions come from boundary states which determine a Regge geometry. By contrast, most non-Regge geometries have suppressed asymptotics. 
 
In the case of a Regge state associated to a non-degenerate 4-simplex boundary geometry, the asymptotic formula contains the cosine of the Regge action 
\be
S_{\mbox{{\tiny Regge}}} = \sum_{a<b} A_{ab} \Theta_{ab}, \nn
\ee
with the areas of the triangles given by $\gamma k_{ab}$, as calculated in the original EPRL paper. Note that it is also logically possible to identify the areas with the spins $k_{ab}$, in which case the Regge action appears in the argument of the complex exponential terms. Interestingly, these terms scale with the same exponent. 

We have also studied the semi-classical limit of the model for more singular boundary data that we have labelled vector geometries. For this type of boundary, the asymptotic formula is not dominated by non-degenerate four-simplex geometries and does not contain the Regge action. For fully non-Regge boundaries, the amplitude is  suppressed in the large spin limit.

An interesting corollary of our work is the asymptotic analysis of the EPR model where the Immirzi parameter is set to zero. In this case, the left and right spins are equal $j_{ab}^+ = j_{ab}^- = j_{ab}$ and our result leads to the following asymptotic formula for a Regge-like boundary state $\psi_\lambda$
\bea
&& f_4(\psi_\lambda)
\sim
(-1)^{\chi'}
\left(\frac{2 \pi}{\lambda}\right)^{12}  \frac{2^8}{ (4\pi)^{16}}
 \nn \\ 
&& \times \left[ 2 N^{\gamma=0}_{+-}   
+ N^{\gamma=0}_{++}\exp{ \left( i
\lambda\sum_{a<b}j_{ab}\Theta_{ab}\right)} +
N^{\gamma=0}_{--}\exp{ \left(-i
\lambda\sum_{a<b}j_{ab}\Theta_{ab}\right)} \right]. \nn
\eea

The completion of this work required an appropriate posing of the asymptotic problem. This has led us to unravel the crucial role of the phase information contained in coherent states. We have indeed shown that, while the phase of coherent states is unimportant in the description of a single coherent tetrahedron, it plays an important geometrical role in the description of more complicated triangulated three-manifolds. We have discovered that the phase of coherent states registers the gluing properties of three-dimensional tetrahedra. Hence, the framework underlying our semi-classical analysis has a flavor of the area/angle formulation of Regge calculus by Dittrich and Speziale \cite{Dittrich:2008va}.

Several extensions of this work require attention. Firstly, the asymptotics for the Lorentzian models \cite{Engle:2007wy,Pereira:2007nh} should be computed. Secondly, a numerical test of the asymptotic formula in comparison to direct calculation of the spin network vertex amplitudes is desirable. Finally, a study of the renormalisation of the model and its behaviour under Pachner moves would be an interesting avenue of investigation.

\section*{Acknowledgements}
WF is supported by the Royal Commission for the Exhibition of
1851.  RD and FH are funded by EPSRC doctoral grants. WF thanks the QG research networking programme of the ESF for a visit grant.


\end{document}